\documentclass[10pt, twocolumn, twoside]{IEEEtran}

\usepackage{amsthm,amsmath,amssymb,amsfonts,mathrsfs,mathtools,dsfont,mathrsfs,bm,bbm}
\DeclareMathAlphabet{\mathpzc}{OT1}{pzc}{m}{it}
\DeclareSymbolFontAlphabet{\amsmathbb}{AMSb}
\usepackage[Symbol]{upgreek}

\usepackage{cite}
\makeatletter
\newcommand{\removelatexerror}{\let\@latex@error\@gobble}
\let\NAT@parse\undefined
\makeatother
\usepackage[hyphens]{url}
\usepackage{hyperref}
\hypersetup{breaklinks, colorlinks, linkcolor=purple, anchorcolor=purple, citecolor=purple, urlcolor=purple}
\usepackage{bookmark}

\usepackage[ruled,vlined,linesnumbered]{algorithm2e}

\usepackage{enumitem}

\usepackage{csvsimple}
\usepackage{tabularx}
\usepackage{makecell}
\newcolumntype{L}{X}
\newcolumntype{C}{>{\centering \arraybackslash}X}
\newcolumntype{R}{>{\raggedright \arraybackslash}X}
\usepackage{booktabs}
\usepackage{multirow}

\usepackage[caption=false]{subfig}
\usepackage{float}
\usepackage{graphicx}
\usepackage{epsfig}
\usepackage{psfrag}
\usepackage{xcolor}

\usepackage{tikz}
\usetikzlibrary{decorations.pathreplacing}
\usetikzlibrary{calc}
\usepackage{forest}
\usepackage{tikz-qtree}
\usepackage{pdflscape} 
\usetikzlibrary{trees} 
\usetikzlibrary{positioning, shapes.geometric}

\definecolor{teal}{rgb}{0.0, 0.5, 0.5}
\definecolor{antiquewhite}{rgb}{0.9803, 0.9215, 0.8431}
\definecolor{magenta}{rgb}{1, 0, 1}
\definecolor{ashgrey}{rgb}{0.7, 0.75, 0.71}
\definecolor{atomictangerine}{rgb}{1.0, 0.6, 0.4}

\usepackage[todonotes={textsize=tiny}]{changes}
\definechangesauthor[name={Ye Guo}, color=orange]{yg}
\definechangesauthor[name={Haitian}, color=teal]{ht}
\newcommand{\beq}{\begin{equation}}
\newcommand{\eeq}{\end{equation}}
\newcommand{\beqa}{\begin{eqnarray}}
\newcommand{\eeqa}{\end{eqnarray}}
\newcommand{\beqan}{\begin{eqnarray*}}
\newcommand{\eeqan}{\end{eqnarray*}}

\newcommand{\rank}{\text{rank }}

\newcommand\T{{\mathpalette\raiseT\intercal}}
\newcommand\raiseT[2]{\raisebox{0.25ex}{$#1#2$}
}

\newcommand{\Dset}{\mathds{D}}

\newcommand{\Rset}{\mathds{R}}

\newcommand{\Yset}{\mathds{Y}}
\newcommand{\Zset}{\mathds{Z}}

\newcommand{\Acal}{{\cal A}}
\newcommand{\Bcal}{{\cal B}}

\newcommand{\Dcal}{{\cal D}}

\newcommand{\Ical}{{\cal I}}

\newcommand{\Lcal}{{\cal L}}
\newcommand{\Mcal}{{\cal M}}
\newcommand{\Ncal}{{\cal N}}

\newcommand{\Pcal}{{\cal P}}

\newcommand{\Scal}{{\cal S}}

\newcommand{\Ucal}{{\cal U}}

\newcommand{\Wcal}{{\cal W}}

\newcommand{\Zcal}{{\cal Z}}

\newcommand{\JJ}{{\bf J}}

\newcommand{\Bsf}{{\sf B}}

\newcommand{\Nsf}{{\sf N}}

\newcommand{\Psf}{{\sf P}}

\newcommand{\Vsf}{{\sf V}}

\newcommand{\bone}{\mathbf{1}}

\renewcommand{\v}[1]{{\bm{#1}}}

\renewcommand{\[}{\left[}
\renewcommand{\]}{\right]}

\renewcommand{\(}{\left(}
\renewcommand{\)}{\right)}

\newcounter{l1}
\newcounter{l2}
\newcounter{l3}
\newcommand{\bdotlist}{\begin{list}{$\bullet$}{}}
\newcommand{\bboxlist}{\begin{list}{$\Box$}{}}
\newcommand{\bbboxlist}{\begin{list}{\raisebox{.005in}{{\tiny
$\blacksquare$ \ \ }}}{}}
\newcommand{\bdashlist}{\begin{list}{$-$}{} }
\newcommand{\blist}{\begin{list}{}{} }
\newcommand{\barablist}{\begin{list}{\arabic{l1}}{\usecounter{l1}}}
\newcommand{\balphlist}{\begin{list}{(\alph{l2})}{\usecounter{l2}}}
\newcommand{\bAlphlist}{\begin{list}{\Alph{l2}.}{\usecounter{l2}}}
\newcommand{\bdiamlist}{\begin{list}{$\diamond$}{}}
\newcommand{\bromalist}{\begin{list}{(\roman{l3})}{\usecounter{l3}}}

\newtheorem{theorem}{Theorem}
\newtheorem{lemma}{Lemma}

\newtheorem{corollary}{Corollary}

\newtheorem{definition}{Definition}

\renewcommand{\bone}{{\mathds{1}}}
\newcommand{\nbd}{{\sf nbd}}

\renewcommand{\rank}{\mathop{\bf rank}}

\newcommand{\cf}{{\it cf.}}

\newcommand{\ie}{{\it i.e.}}
\newcommand{\etc}{{\it etc.}}

\newcommand{\wrt}{{\it w.r.t. }}
\newcommand{\etal}{{\it et al. }}

\newcommand{\CR}{{\sf CR}}
\newcommand{\FC}{{\sf FC}}
\definecolor{light-gray}{gray}{0.95}
\newcommand{\code}[1]{\colorbox{light-gray}{\texttt{#1}}}

\newcommand{\mpLCP}{{\sf mpLCP}}
\newcommand{\thetafix}{\tilde{\v{\theta}}}
\newcommand{\thetaproj}{\bar{\v{\theta}}}
\newcommand{\thetatmp}{\hat{\v{\theta}}}
\newcommand{\thetaopt}{\v{\theta}^\star}
\newcommand{\thetak}{\v{\theta}^k}

\renewcommand{\int}{\mathop{\bf int}}

\begin{document}

\title{\Large{\bf{
  On Degeneracy Issues in Multi-parametric Programming and Critical Region Exploration based Distributed Optimization 
    in Smart Grid Operations
}}}

\author{
Haitian Liu, \textit{Student Member}, \textit{IEEE},
Ye Guo, \textit{Senior Member}, \textit{IEEE}, 
Hao Liu, \textit{Student Member}, \textit{IEEE}
\thanks{
Haitian Liu, Ye Guo are with Tsinghua-Berkeley Shenzhen Institute (TBSI), Tsinghua University, Shenzhen, Guangdong 518055, China. 
Hao Liu is with China Electric Power Planning \& Engineering Institute, Beijing 100120, China.
This work is supported in part by the National Natural Science Foundation of China under Grant 51977115.

Corresponding author: Ye Guo, e-mail: \url{guo-ye@sz.tsinghua.edu.cn}.}}

\newcounter{MYtempeqncnt}

\maketitle


\begin{abstract}
    Improving renewable energy resource utilization efficiency is crucial to reducing carbon emissions, and multi-parametric programming has provided a systematic perspective in conducting analysis and optimization toward this goal in smart grid operations.
  This paper focuses on two aspects of interest related to multi-parametric linear/quadratic programming (mpLP/QP). 
  First, we study degeneracy issues of mpLP/QP. 
  A novel approach to deal with degeneracies is proposed to find all critical regions containing the given parameter.
  Our method leverages properties of the multi-parametric linear complementary problem, vertex searching technique, and complementary basis enumeration. 
  Second, an improved critical region exploration (CRE) method to solve distributed LP/QP is proposed under a general mpLP/QP-based formulation. 
  The improved CRE incorporates the proposed approach to handle degeneracies. 
  A cutting plane update and an adaptive stepsize scheme are also integrated to accelerate convergence under different problem settings.
  The computational efficiency is verified on multi-area tie-line scheduling problems with various testing benchmarks and initial states.  
\end{abstract}


\section{Introduction}


Multi-parametric programming (mpP) systematically studies variations of optimal solutions concerning a set of parameters of interest. 
The mpLP/QP can adapt to many practical problems, such as explicit model predictive control \cite{borrelli_bemporad_morari_2017,novoselnikParametricOptimizationBased2020a} and process system engineering \cite{pappasMultiparametricProgrammingProcess2021,akbariImprovedMultiparametricProgramming2018a}. 
The modern power system is enduring an increasing penetration of renewable energy resources (RESs) for cleaner generation. 
Albeit enjoying a near-zero carbon footprint, RESs have highly unbalanced geographical distributions. 
Enhancing a deeper combination of mpP-based techniques and RES utilization is crucial to improving overall efficiency in smart grid operations.
In power system fields, parametric solutions of mpLP/QP have been applied to conduct congestion management \cite{jiProbabilisticForecastingRealTime2017}, estimate distributed energy resources (DERs) hosting capacity in an active distribution network \cite{taheriFastProbabilisticHosting2021} as well as facilitate the corresponding peer-to-peer energy sharing mechanism \cite{chenFlexibilityRequirementWhen2022}, and solve tie-line scheduling problems in large-scale interconnected power networks \cite{guoCoordinatedMultiAreaEconomic2017, guoRobustTieLineScheduling2018}. 
Due to the complexity of topologies and constituting components, degeneracies broadly exist when we try to characterize the parametric relations within the power system. 
Such an issue has a latent risk that the existing mpLP/QP algorithm may not be able to proceed normally.

A direct sign of degeneracy for multi-area tie-line scheduling problems is nonunique generation schedules or nonunique locational marginal prices. 
Theoretically, the degeneracy of an mpLP/QP indicates the optimal solutions under given parameters either are nonunique or violate the strict complementary slackness condition. 
An mpLP/QP can be solved by geometric \cite{borrelliGeometricAlgorithmMultiparametric2003} and combinatorial \cite{guptaNovelApproachMultiparametric2011a} based approaches in the state-of-the-art research.
Both methods are based on the ideas of the active set method, where the constraint sets are divided into active and inactive sets.
As a result, the parameter space is partitioned into a group of polyhedral critical regions ($\CR$). 
Each $\CR$ corresponds to a range of parameters where the active constraint indices remain unchanged. 
The geometric mpLP/QP is based on $\CR$'s graphical adjacency to explore the parameter space explicitly. 
In contrast, the combinatorial ones partition the parameter space implicitly by enumerating the active set candidate lists of the constraint indices \cite{ahmadi-moshkenaniCombinatorialApproachMultiparametric2018}. 
The former is vulnerable to degeneracy but scales nicely with a large-scale system. The latter is less vulnerable to degeneracy, but the scalability is unsatisfactory with the growing number of variables and constraints due to the exhaustive enumeration. 

The general procedure of the geometric-based approach consists of two basic steps \cite{guptaNovelApproachMultiparametric2011a}:
(a) For a given parameter, determine the optimal solution as a parameter-dependent function, valid over a specific $\CR$. 
(b) Explore the remaining parameter space by searching all adjacent $\CR$s of the current $\CR$. 
Degeneracy may occur in step (a). 
When it happens, the $\CR$ containing the given parameter may not be full-dimensional or uniquely defined. 
Under such a case, it is hard to identify adjacent $\CR$s, and step (b) cannot proceed normally. 

To address the degeneracy issue, existing works apply three strategies:
One approach is randomly perturbing the given parameter in step (a) to see whether a nondegenerate parameter-dependent function exists in its neighborhood \cite{borrelliGeometricAlgorithmMultiparametric2003}. 
The advantage is it directly avoids some degenerate situations, especially when characterizing the sensitivity of the current parameter is inevitable. 
The disadvantage is that if the given parameter is within a full-dimensional $\CR$, then any perturbation cannot resolve degeneracy.
Another approach is to apply generalized inverse \cite[Sec. 6.2.2, 6.3.2]{borrelli_bemporad_morari_2017} \cite{akbariImprovedMultiparametricProgramming2018a} or orthogonal projection techniques \cite{tondelFurtherResultsMultiparametric2003,jonesPolyhedralProjectionParametric2008} for the given parameter when degeneracy happens. 
However, the resulting $\CR$ is generally low-dimensional or even reduced to a singleton. 
Besides, the projection operation may be too costly, and the sensitivities concerning the change of the optimal cost may only be an underestimation. 
The final approach is to solve auxiliary problems \cite{akbariImprovedMultiparametricProgramming2018a,spjotvoldMethodObtainingContinuous2005, spjotvoldContinuousSelectionUnique2007} or apply predefined rules to perturb the original problem \cite{jonesLexicographicPerturbationMultiparametric2007}.
In this way, certain optimal solutions can be obtained depending on the practitioner's settings when there is degeneracy.
The advantage is that we may obtain a full-dimensional $\CR$ along with parametric relations defined over it.
Albeit only a partial characterization of the sensitivities for the given parameter is revealed under such schemes, it is generally enough to explore the remaining parameter space.
However, the disadvantage is that we may require distinct auxiliary objectives or rules for different problems and degeneracy types, which may be complex to design. 

Although the combinatorial-based approach proposed in \cite{ahmadi-moshkenaniCombinatorialApproachMultiparametric2018} and \cite{oberdieckExplicitModelPredictive2017a} shows some possible schemes that degeneracies can be resolved by enumeration. 
Yet the discussions of such schemes are currently still limited to primal degeneracies of relative small-scale strictly convex mpQP problems.
There is still a need for a robust and efficient approach to handle degeneracy for general mpLP/QP, both in engineering practices and algorithm developments.

It is worth noting that the multi-parametric linear complementary problem (mpLCP) has recently attracted research interest, which can be viewed as a generalized form of mpLP/QP. 
Following the same path, there are variants for the geometric \cite{jonesMultiparametricLinearComplementarity2006a,adelgrenTwophaseAlgorithmMultiparametric2016} and combinatorial \cite{hercegEnumerationbasedApproachSolving2015a} based mpLCP approaches.
Notably, all variables in mpLP/QP are transformed into complementary solution pairs in mpLCP. Hence, there is no need to distinguish problem formulations or degeneracy types. 
However, when there are degeneracies in mpLCP, existing works still apply the techniques in mpLP/QP that we discussed above to resolve them. 
Such schemes make the current mpLCP-based algorithms still suffer from the same limitations when degeneracies exist.
In this regard, we proposed a novel degeneracy handling method that leverages both features of geometric- and combinatorial-based algorithms to resolve various degenerate situations in a unified way. 
The proposed scheme can grasp the properties in mpLCP, and all the $\CR$s containing the given degenerate parameter can be obtained efficiently.
Hence, it suits large-scale power systems' robust operation with fast computational requirements.

Due to privacy concerns and computation ability limitations, a distributed solution technique is also preferred in smart grid to improve RESs operational efficiency. 
Such methods solve the problem by optimizing the local dispatch and updating the system boundary state iteratively until convergence. 
Distributed optimization technique is of crucial importance in multiple power system applications.
Currently, researches on solving tie-line scheduling problem primarily focus on dual decomposition-based techniques, which are known to have convergence issues when the system scales, see a recent survey in \cite{kargarianDistributedDecentralizedDC2018}. The primal decomposition-based critical region exploration (CRE) \cite{guoCoordinatedMultiAreaEconomic2017, guoRobustTieLineScheduling2018} method shows fast finite convergent property on many tie-line scheduling problems. However, realizing such an efficient scheme relies on deriving the parametric relation between the local economic dispatch problem and boundary phase angles. Degeneracy might be inevitable when multiple generators have similar generation costs or the system topologies become complex. Such a latent risk prevents the further application of CRE. Similar implementation challenges are also exposed to coordinating distributed energy resources via cloud computing platforms \cite{wangCloudComputingLocal2020}, integrated transmission \& active distribution networks economic dispatch \cite{linDecentralizedDynamicEconomic2018} in obtaining the relevant parametric relations. Hence, the proposed degeneracy handling scheme can greatly complement mpP-based distributed techniques such as CRE. And improving CRE-like schemes can enable efficient coordination for various grid optimization examples among system operators.

The remainder of this paper is organized as follows.  Section \ref{sec:Deg_LC} introduces a novel searching technique to deal with degeneracies, and section \ref{sec:improved_CRE} incorporates the proposed method into CRE and designs an improved scheme. Section \ref{sec:case_new} verifies our improvements under various tie-line scheduling testing systems. Finally, section \ref{sec:conclusion} concludes the paper.

\section{Dealing with degeneracies with multi-parametric linear complementary problems}
\label{sec:Deg_LC}

\subsection{Muiti-parametric linear/quadratic programming problem}

Consider a general convex multi-parametric quadratic programming problem with variables $\v{x} \in \Rset^{n}$ and parametrized in $\v{\theta} \in \Rset^{d}$ \cite[Eq. (6)]{hercegEnumerationbasedApproachSolving2015a}:
\begin{subequations}
    \label{eq:mpSemiQP}
    \begin{alignat}{3}
    \JJ(\v{\theta}) := \ 
    & \underset{\v{x} \in \Rset^{n}}{\text{minimize}} 
    && \ \frac{1}{2} \v{x}^\T \v{H} \v{x} + \v{f}^\T \v{x}, 
    && 
    \label{eq:mpSemiQP.obj} \\ 
    & \text{subject to}
    && \ \v{A} \v{x} \leq \v{b} + \v{C} \v{\theta},
    && : \v{\lambda} \in \Rset^{m}.
    \label{eq:mpSemiQP.cons}
    \end{alignat}
\end{subequations}

Here, the objective \eqref{eq:mpSemiQP.obj} is a convex function with $\v{H} \succeq 0$. 
Variables $\v{x}$ are restricted in constraint \eqref{eq:mpSemiQP.cons}, where the right-hand-side of \eqref{eq:mpSemiQP.cons} is parametrized linearly in $\v{\theta}$.
The dual variables corresponding to \eqref{eq:mpSemiQP.cons} are denoted by $\v{\lambda}$.
Notation $\JJ(\v{\theta})$ represents the value function of problem \eqref{eq:mpSemiQP}, which depicts how optimal cost changes with the parameter $\v{\theta}$. 
Our work also entails two special multi-parametric formulations by nature, \ie, an mpLP when $\v{H} = 0$ \cite[Eq. (1)]{borrelliGeometricAlgorithmMultiparametric2003}, and a strictly convex mpQP when $\v{H} \succ 0$ \cite[Eq. (2)]{guptaNovelApproachMultiparametric2011a}.

Let $\thetafix$ be a given parameter, $(\v{x}(\thetafix), \v{\lambda}(\thetafix))$ is an optimal solution pair to \eqref{eq:mpSemiQP}. 
Then the corresponding optimal active and inactive sets $\Acal$, $\Ical$ are characterized by
\begin{gather}
    {
    \begin{gathered}
        \Acal := \{ j \in \Mcal \ \vert \ \v{A}^{j} \v{x}(\thetafix) - b^{j} - \v{C}^{j} \thetafix = 0 \}, \\
        \Ical := \{ j \in \Mcal \ \vert \ \v{A}^{j} \v{x}(\thetafix) - b^{j} - \v{C}^{j} \thetafix < 0 \},
    \label{eq:set_partition}
    \end{gathered}
    }
\end{gather}
where $\Mcal = \{ 1, \ldots, m \}$ is the index set to constraint \eqref{eq:mpSemiQP.cons}.
Under $\Acal$, $\Ical$, Karush–Kuhn–Tucker (KKT) condition implies
\begin{subequations}
    \label{eq:mpP_KKT}
    \begin{alignat}{2}
        &&& 
        \begin{bmatrix}
            \v{H} & \v{A}^{\Acal, \T}\\
            -\v{A}^{\Acal} & \v{0}
        \end{bmatrix} \begin{bmatrix}
            \v{x}(\thetafix) \\
            \v{\lambda}^{\Acal}(\thetafix)
        \end{bmatrix} = - \begin{bmatrix}
            \v{f} \\
            \v{b}^{\Acal}
        \end{bmatrix} - \begin{bmatrix}
            \v{0}\\
            \v{C}^{\Acal}
        \end{bmatrix} \thetafix,
        \label{eq:mpP_KKT.eq} \\
        &&& 
        \v{A}^{\Ical} \v{x}(\thetafix) - b^{\Ical} - \v{C}^{\Ical} < \v{0}, 
        \quad \v{\lambda}^{\Acal}(\thetafix) \geq \v{0}.
        \label{eq:mpP_KKT.ieq} 
    \end{alignat}
\end{subequations}

When the problem \eqref{eq:mpSemiQP} is nondegenerate under $\thetafix$, the coefficient matrix on the left-hand-side of \eqref{eq:mpP_KKT.eq} is invertible.
The parametric mapping of primal-dual solution pair is written as 
\begin{alignat}{2}
    \begin{bmatrix}
        \v{x}(\v{\theta}, \thetafix) \\
        \v{\lambda}^\Acal(\v{\theta}, \thetafix)
    \end{bmatrix} 
    & = - \begin{bmatrix}
            \v{H} & \v{A}^{\Acal, \T}\\
            -\v{A}^{\Acal} & \v{0}
        \end{bmatrix}^{-1} \left( \begin{bmatrix}
        \v{f} \\
        \v{b}^{\Acal}
    \end{bmatrix} + \begin{bmatrix}
        \v{0}\\
        \v{C}^{\Acal}
    \end{bmatrix} \v{\theta} \right) \notag \\
    & := \begin{bmatrix}
        \v{T}_x \\
        \v{T}_{\Acal}
    \end{bmatrix} \v{\theta} + \begin{bmatrix}
        \v{k}_x \\
        \v{k}_{\Acal}
    \end{bmatrix}, \quad \v{\lambda}^\Ical(\v{\theta}, \thetafix) := \v{0}.
    \label{eq:KKT_optimizer}
\end{alignat}

Equation \eqref{eq:KKT_optimizer} is derived from \eqref{eq:mpP_KKT.eq} and is also based on the fact that dual variables for inactive constraints are equal to zero.
Each $\CR$ characterizes an optimal partition to the set of active/inactive constraint indices \eqref{eq:mpSemiQP.cons}. 
It is derived by substituting \eqref{eq:KKT_optimizer} into \eqref{eq:mpP_KKT.ieq}, indicating the region of $\v{\theta}$ where the partition $\{\Acal, \Ical\}$ does not change.
\begin{gather}
    { 
    \begin{gathered}
        \CR[\thetafix] := \{ \v{\theta} \ \vert \ \hat{\v{D}} \v{\theta} \leq \hat{\v{r}} \}, \\
        \hat{\v{D}} = \begin{bmatrix}
            \v{A}^{\Ical} \v{T}_x - \v{C}^{\Ical} \\
            -\v{T}_{\Acal}
        \end{bmatrix}, \quad 
        \hat{\v{r}} = \begin{bmatrix}
            \v{b}^{\Ical} - \v{A}^{\Ical} \v{k}_x \\
            \v{k}_{\Acal}
        \end{bmatrix}.
        \label{eq:KKT_CR}
    \end{gathered}
    }
\end{gather}

By substituting \eqref{eq:KKT_optimizer} into \eqref{eq:mpSemiQP.obj}, the value function segment $\JJ(\v{\theta}, \thetafix)$ defined over $\CR[\thetafix]$ is
\begin{gather}
    { 
    \begin{gathered}
        \JJ(\v{\theta}, \thetafix) := 
        \frac{1}{2} \v{\theta}^\T \hat{\v{H}} \v{\theta} 
        + \hat{\v{f}}^\T \v{\theta} + \hat{c}, 
        \label{eq:KKT_VF} \\
        \hat{\v{H}} = \v{T}_x^\T \v{H} \v{T}_x, \quad 
        \hat{\v{f}} = \v{T}_x^\T \v{H}^\T \v{k}_x + \v{T}_x^\T \v{f}, \\
        \hat{c} = \frac{1}{2} \v{k}_x^\T \v{H} \v{k}_x + \v{f}^\T \v{k}_x.
    \end{gathered}
    }
\end{gather}

\subsection{Characterizing different types of degeneracies}

If an mpLP/QP is feasible under a given parameter, then there at least exists a solution that satisfies the KKT condition. The returned solution is said to be nondegenerate given
\begin{itemize}[leftmargin=*]
    \item linear independence constraint qualification (LICQ),
    \item second-order sufficient condition (SOSC),
    \item strict complementarity slackness (SCS),
\end{itemize}
are all satisfied \cite[Th. 3.1]{fiaccoDegeneracyNLPDevelopment1993}. Violation of any of the aforementioned conditions may cause two abnormal situations.
\begin{enumerate}[leftmargin=*]
    \item The constraint partition in \eqref{eq:set_partition} may lead to singularity to the coefficient matrix in the left-hand-side of \eqref{eq:mpP_KKT.eq}. This applies to LICQ or SOSC violations,
    \item The coefficient matrix is nonsingular. However, some dual variables regarding the active constraints in \eqref{eq:KKT_optimizer} are zeros. This applies to the SCS violations.  
\end{enumerate}

\begin{figure}[htbp]
    \centering 
    \vspace{-15pt}
    \subfloat[]{
        \includegraphics[width=0.235\textwidth]{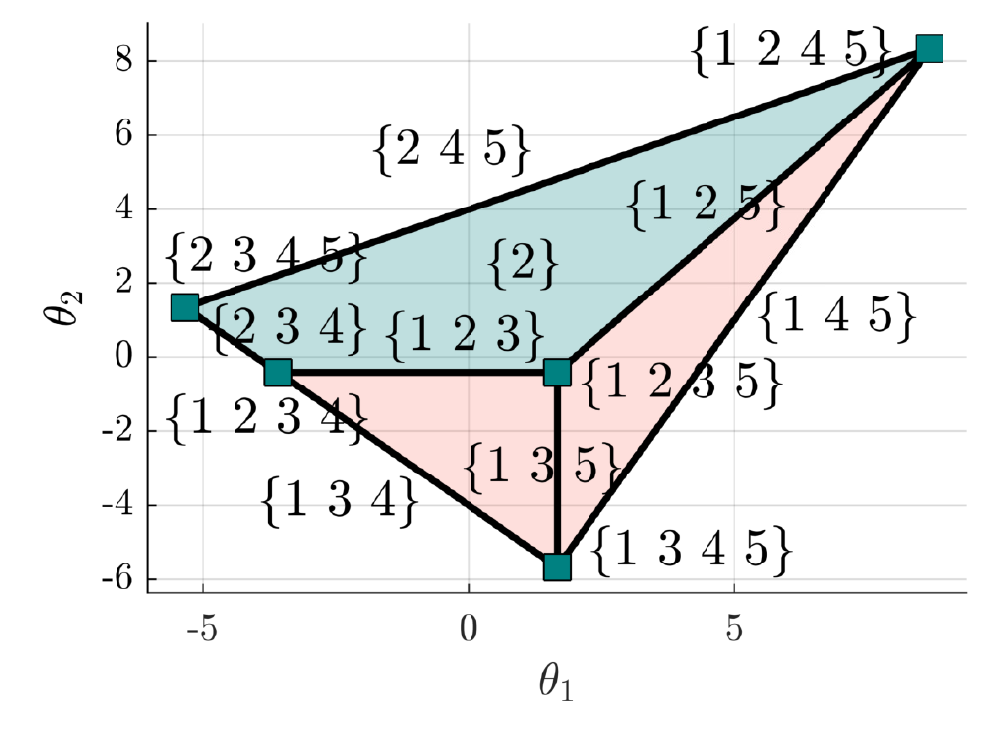}
        \label{fig:deg.woSOSC_LICQ}
    }
    \subfloat[]{
        \includegraphics[width=0.235\textwidth]{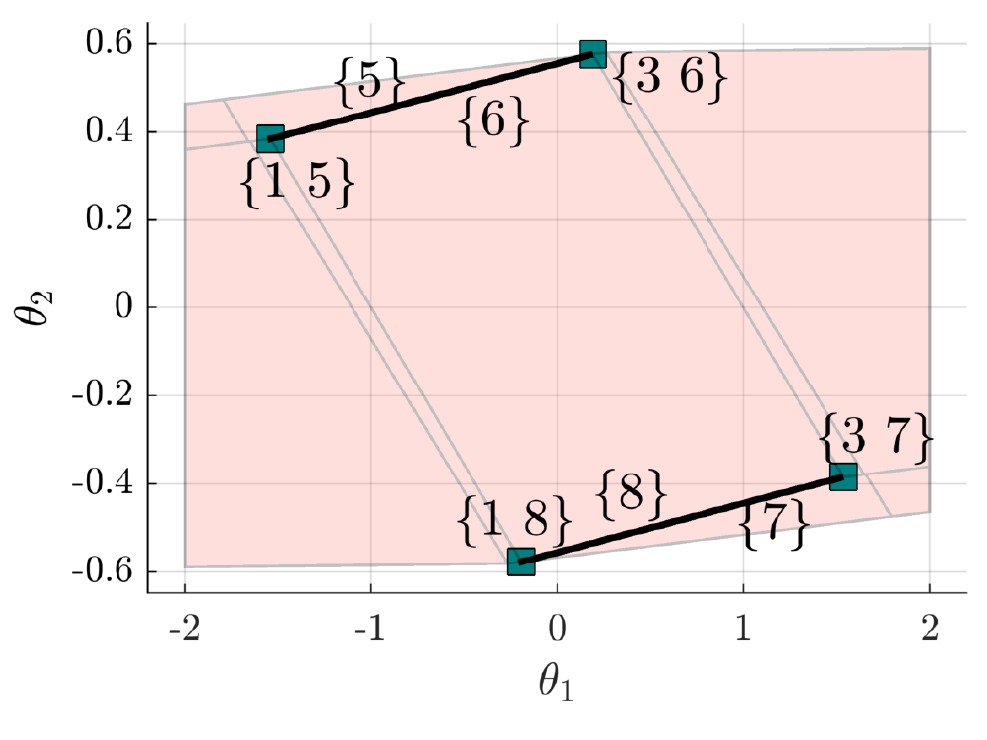}
        \label{fig:deg.woSCS}
    }          
    \vspace{-10pt}
    \caption{
    Illustrations on the violation of 
    (a) LICQ/SOSC, 
    (b) SCS condition.
    Indices in the figures represent the active constraints set, and the degenerated region is shown in a green-shaded area, thick lines, and dots. Red shaded areas represent the non-degenerate region of the problems.
    Due to the singularity, the degenerate space is plotted by directly projecting the variable space onto the parameter space. 
    }
    \label{fig:deg}
    \vspace{-7pt}
\end{figure}

The first case is illustrated in Fig. \ref{fig:deg.woSOSC_LICQ}, where the data is taken from \cite[Eq. (6.32)-(6.33)]{borrelli_bemporad_morari_2017}. 
For the second case, an example taken from \cite{tondelAlgorithmMultiparametricQuadratic2003a} is shown in Fig. \ref{fig:deg.woSCS}.

Current degeneracy types can be categorized into primal and dual degeneracies.
The former implies LICQ violations and has nonunique dual optimal solutions under the given parameter.
The latter violates SOSC and has nonunique primal optimal solutions. 
Violation of the SCS condition can be viewed as a transition state among full-dimensional critical regions that are geometrically adjacent. Moreover, the same critical region may not be uniquely defined, as shown in Fig. \ref{fig:deg.woSCS}. When both primal and dual degeneracies exist, current approaches are hard to identify and handle, making mpLP/QP-based algorithms unable to proceed efficiently.

\subsection{Find all critical regions containing the given parameter}

The given parameter $\thetafix$ may result in degeneracies if it belongs to degenerate regions as indicated in Fig. \ref{fig:deg}.
In subsequent contents, we propose a unified approach to deal with degeneracies. 
The proposed approach is mainly inspired by the previous works \cite{parisMultipleOptimalSolutions1983a, kanekoNumberSolutionsClass1979, hercegEnumerationbasedApproachSolving2015a}. Their efforts and our contributions are listed as follows.
\begin{enumerate}[leftmargin=*]
    \item Paris \etal \cite{parisMultipleOptimalSolutions1983a} provides a framework such that the nonuniqueness of QP can be identified under \cite{kanekoNumberSolutionsClass1979} via LCP and all solutions can be found after reformulation. 
    Our work extends such a case from QP to mpQP and analyzes the degenerate parametric relation under mpLCP.
    \item Herceg \etal \cite{hercegEnumerationbasedApproachSolving2015a} proposed a combinatorial-based approach to solve the mpP problem by exhaustive enumeration. Our work applies an efficient partial enumeration strategy to resolve degeneracy for certain parameters of interest.
\end{enumerate}
The procedure consists of steps as listed below. The logical relation of these steps is synthesized in Algo. \ref{alg:LC-Pivoting}.

{\noindent (1) Transform mpLP/QP \eqref{eq:mpSemiQP} to mpLCP \eqref{eq:mpLCP_fin},}

{\noindent (2) Solve the mpLCP \eqref{eq:mpLCP_fin} under $\thetafix$,}

{\noindent (3) Verify the uniqueness of mpLCP \eqref{eq:mpLCP_fin}’s solution,}

{\noindent (3-1) If unique, save the current complementary solution,}

{\noindent (3-2) If not, save the vertices of mpLCP \eqref{eq:mpLCP_fin}’s solution set,}

{\noindent (4) Enumerate complementary bases to the set's all vertices,}

{\noindent (5) Obtain all critical regions containing $\thetafix$ using all complementary bases.}

Step (1): transform mpLP/QP \eqref{eq:mpSemiQP} to mpLCP \eqref{eq:mpLCP_fin}.
To start with, we slightly reformulate the original mpLP/QP \eqref{eq:mpSemiQP} into the following equivalent form
\begin{subequations}
    \label{eq:mpLCP_ini}
    \begin{alignat}{3}
    \JJ(\v{\theta}) := \ 
    & \underset{\v{x} \in \Rset^n}{\text{minimize}} 
    && \ \frac{1}{2} \v{x}^\T \v{H} \v{x} + \v{f}^\T \v{x}, 
    && 
    \label{eq:mpLCP_ini.obj} 
    \\ 
    & \text{subject to}
    && \ \v{A} \v{x} + \v{s} = \v{b} + \v{C} \v{\theta},
    && : \v{\lambda} \in \Rset^m,
    \label{eq:mpLCP_ini.cons1} 
    \\
    &
    && \ \v{x} \geq \v{0},
    && : \v{\mu} \in \Rset^n.
    \label{eq:mpLCP_ini.cons2}
    \end{alignat}
\end{subequations}

Here, we slightly abuse the notation and dimension to maintain the consistency of the description. 
Notations $\v{\lambda}$ and $\v{\mu}$ are the dual variables for constraints \eqref{eq:mpLCP_ini.cons1}-\eqref{eq:mpLCP_ini.cons2}. 
The KKT conditions of \eqref{eq:mpLCP_ini} can be written as 
\begin{alignat*}{2}
    \v{H} \v{x} + \v{f} + \v{A}^\T \v{\lambda} - \v{\mu} = \v{0}, 
    & \ \v{\lambda}, \v{\mu} \geq \v{0}, \\
    \mu_j x_j = 0, 
    & \ \forall \ j = 1, \ldots, n, \\ 
    \lambda_j s_j = 0, 
    & \ \forall \ j = 1, \ldots, m, \\
    \v{A} \v{x} + \v{s} = \v{b} + \v{C} \v{\theta}, 
    & \ \v{s}, \v{x} \geq \v{0}.
\end{alignat*}
By rearranging the terms above, we have
\begin{subequations}
    \begin{alignat}{2}
    &&& \begin{bmatrix}
        \v{\mu} \\
        \v{s}
    \end{bmatrix} = \begin{bmatrix}
        \v{H} & \v{A}^\T \\
        -\v{A} & \v{0}
    \end{bmatrix} \begin{bmatrix}
        \v{x} \\
        \v{\lambda}
    \end{bmatrix} + \begin{bmatrix}
        \v{f} \\
        \v{b}
    \end{bmatrix} + \begin{bmatrix}
        \v{0}\\
        \v{C}
    \end{bmatrix} \v{\theta}, \\
    &&& \v{s}^\T \v{\lambda} = 0, \ \v{\mu}^\T \v{x} = 0, \ \v{s}, \v{\lambda}, \v{\mu}, \v{x} \geq \v{0}.
    \end{alignat}
    \label{eq:mpLCP_mtx}
\end{subequations}
Define
\begin{gather}
{ 
\begin{gathered}
    \v{M} = \begin{bmatrix}
        \v{H} & \v{A}^\T \\
        -\v{A} & \v{0}
    \end{bmatrix}, \quad 
    \v{q} = \begin{bmatrix}
        \v{f} \\
        \v{b}
    \end{bmatrix}, \quad 
    \v{Q} = \begin{bmatrix}
        \v{0}\\
        \v{C}
    \end{bmatrix}, \\
    \v{w} = \begin{bmatrix}
        \v{\mu}^\T &
        \v{s}^\T
    \end{bmatrix}^\T, \quad 
    \v{z} = \begin{bmatrix}
        \v{x}^\T &
        \v{\lambda}^\T
    \end{bmatrix}^\T.
    \label{eq:mpLCP_coef}
\end{gathered}
}
\end{gather}
The mpLCP can be written as
\begin{subequations}
    \label{eq:mpLCP_fin}
    \begin{alignat}{2}
    &&\text{find} & \quad \v{w}, \v{z}, \\
    &&\text{subject to} & \quad \v{w} = \v{M} \v{z} + \v{q} + \v{Q} \v{\theta}, \label{eq:mpLCP_fin.1} \\
    &&& \quad \v{w}^\T \v{z} = 0, \quad \v{w}, \v{z} \geq \v{0}, \label{eq:mpLCP_fin.2}
    \end{alignat}
\end{subequations}
where $\v{w}$, $\v{z} \in \Rset^p$, and $p$ is called the order of the mpLCP. By \eqref{eq:mpLCP_coef}, we know $p = n + m$.
As \eqref{eq:mpLCP_fin.1}-\eqref{eq:mpLCP_fin.2} represents the KKT conditions of \eqref{eq:mpLCP_ini}, any feasible solution of mpLCP \eqref{eq:mpLCP_fin} is an optimal solution to mpLP/QP \eqref{eq:mpLCP_ini}.

By \cite{hercegEnumerationbasedApproachSolving2015a}, let $\v{Y} = \left [ \v{I}, \ -\v{M} \right ]$, $\v{y} = \left [ \v{w}^\T, \ \v{z}^\T \right ]^\T$, we arrive at a more compact form of \eqref{eq:mpLCP_fin} as
\begin{subequations}
    \label{eq:mpLCP_fin.compact}
    \begin{alignat}{2}
    &&\text{find} & \quad \v{y}, \\
    &&\text{subject to} & \quad \v{Y}_{\cdot, \Bcal} \v{y}_{\Bcal} = \v{q} + \v{Q} \v{\theta}, \\
    &&& \quad \v{y}_{\Bcal} \geq \v{0}, \quad \v{y}_{\Ncal} = \v{0},
    \end{alignat}
\end{subequations}
where the index set $\Bcal \subset \left \{ 1, 2, \ldots, 2p \right \}$ is called a complementary basis of mpLCP in \eqref{eq:mpLCP_fin}. 
The cardinality of $\Bcal$ is denoted by $\vert \Bcal \vert$.
The complement of $\Bcal$ is $\Ncal = \left \{ 1, 2, \ldots, 2p \right \} \backslash \Bcal$. 
The matrix $\v{Y}_{\cdot, \Bcal} \in \Rset^{p \times \vert \Bcal \vert} $ is formed from $\v{Y}$ taking all rows and columns indexed by $\Bcal$.
With partition $\Bcal$, $\Ncal$, $\v{y}_{\Bcal}$ and $\v{y}_{\Ncal}$ is called the basic and nonbasic variables, respectively. 
To ensure any $\v{y}$ satisfy compact form \eqref{eq:mpLCP_fin.compact} is a feasible solution to \eqref{eq:mpLCP_fin}, a sufficient condition related to $\Bcal$ is all of the following requirements are satisfied \cite{hercegEnumerationbasedApproachSolving2015a}:
\begin{itemize}[leftmargin=*]
    \item cardinality of set $\Bcal$ satisfies $\vert \Bcal \vert = p$,
    \item rank of matrix $\v{Y}_{\cdot, \Bcal}$ satisfies $\rank(\v{Y}_{\cdot, \Bcal}) = p$,
    \item exact one element of the index pair $(i, p+i)$, $\forall \ i = 1, \ldots, p$, belongs to set $\Bcal$.
\end{itemize}

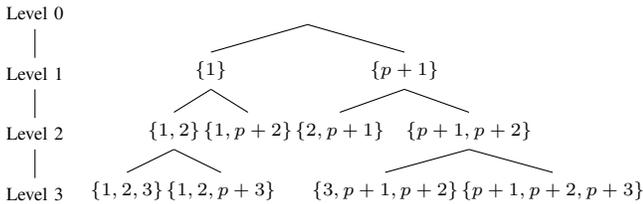
\begin{figure}[htbp]
    \centering         
    \vspace{-10pt}
    \begin{tikzpicture}
        \tikzset{font=\scriptsize, level distance=0.8cm}
        \Tree [.Level\ 0 [.Level\ 1 [.Level\ 2 [.Level\ 3 ] ] ] ]
    \end{tikzpicture}         
    \begin{tikzpicture}  
        \tikzset{font=\scriptsize, 
        level distance=0.8cm,
        level 1/.style={sibling distance=-2mm},
        level 2/.style={sibling distance=-12mm},
        level 3/.style={sibling distance=-2mm}}
        \Tree [.{} 
        [.\{$1$\} 
        [.\{$1,2$\} 
        [.\{$1,2,3$\} ] [.\{$1,2,p+3$\} ] ] 
        [.\{$1,p+2$\} ] ] 
        [.\{$p+1$\} 
        [.\{$2,p+1$\} ] 
        [.\{$p+1,p+2$\} 
        [.\{$3,p+1,p+2$\} ] [.\{$p+1,p+2,p+3$\} ]
        ] ]
        ]              
    \end{tikzpicture}
    \vspace{-7pt}
    \caption{Illustrative process to generate basis candidates \cite[Fig. 1]{hercegEnumerationbasedApproachSolving2015a}. When $p=3$, all possibles candidates are listed in level 3.}
    \label{fig:tree}
\end{figure}

The importance of the formulation \eqref{eq:mpLCP_fin.compact} is that the complementary condition $\v{w}^\T\v{z}=0$ is no longer needed. Instead, it has been encoded concerning the selection of basis $\Bcal$. 
For any given $\thetafix$, if there is a unique basis $\Bcal$ satisfy \eqref{eq:mpLCP_fin.compact}, then it indicates there is a unique feasible solution to \eqref{eq:mpLCP_fin} under the $\thetafix$. 
Hence, solving \eqref{eq:mpLCP_fin} can now be viewed as bases selection procedure. 
One essential step to identifying feasible bases is basis candidates enumeration. There are $2^p$ possible combinations of $\Bcal$ for mpLCP with order $p$. 
We can enumerate all possible complementary bases by selecting the basic variables in ascending order. 
This leads to a tree structure, as shown in Fig. \ref{fig:tree}.
All bases can be obtained at level $p$ for mpLCP with order $p$.

We need to mention that our focus is to deal with degeneracy for some specific parameters. Hence, there is no need to find all feasible bases.
Instead, we are only interested in searching for some of the bases that can generate $\CR$ containing the given $\thetafix$. In this way, we can always characterize the sensitivities in a neighborhood of any $\thetafix$ regardless of degeneracy. 
Note, there is no consideration of utilizing a given $\thetafix$ in existing combinatorial-based mpQP \cite{guptaNovelApproachMultiparametric2011a}, and mpLCP \cite{hercegEnumerationbasedApproachSolving2015a} algorithms.

Step (2): solve the mpLCP \eqref{eq:mpLCP_fin} under the given parameter.
Given $\thetafix$, the mpLCP \eqref{eq:mpLCP_fin} reduces to a linear complementary problem (LCP). 
Such an LCP can be solved by a mature complementary pivot algorithm called Lemke's method \cite[Ch. 2]{Murty1988-lx}. 
Denote $\bar{\Bcal}$ as the basis obtained from Lemke's method and let $\bar{\v{B}} = \v{Y}_{\cdot, \bar{\Bcal}}$ be the corresponding basis matrix. 
The transformation of $\v{M}$, $\v{q}$, $\v{Q}$ in the final tableau of Lemke's method is denoted by $\bar{\v{M}}$, $\bar{\v{q}}$, $\bar{\v{Q}}$ and can be calculated as
\begin{gather}
    { 
    \begin{gathered}
        \bar{\v{M}} = \bar{\v{B}}^{-1}\v{M}, \ 
        \bar{\v{q}} = \bar{\v{B}}^{-1}\v{q}, \
        \bar{\v{Q}} = \bar{\v{B}}^{-1}\v{Q}, \
        \label{eq:mpLCP.Lemke.ini.coef}
    \end{gathered}
    }
\end{gather}
let $\bar{\Ncal}$ be the complement of $\bar{\Bcal}$, the solution to this LCP can be expressed as
\begin{gather}
    { 
    \begin{gathered}
        \bar{\v{y}}_{\bar{\Bcal}}(\thetafix) = \bar{\v{q}} + \bar{\v{Q}} \thetafix, 
        \ \bar{\v{y}}_{\bar{\Ncal}}(\thetafix) = \v{0},
        \label{eq:mpLCP.Lemke.ini.sol}
    \end{gathered}
    }
\end{gather}
and $\bar{\v{w}}(\thetafix)$, $\bar{\v{z}}(\thetafix)$ can be readily obtained via rearranging the elements in $\bar{\v{y}}_{\bar{\Bcal}}(\thetafix)$, $\bar{\v{y}}_{\bar{\Ncal}}(\thetafix)$.
Note that $\bar{\Bcal}$ is selected based on the complementary pivoting rule. Hence, $\bar{\v{B}}$ must be an invertible square matrix of rank $p$\footnote{
One may prefer to solve \eqref{eq:mpSemiQP} with a fixed $\v{\theta}$ directly as a linear/quadratic programming (LP/QP) problem, which can be handled readily by state-of-art solvers. If LP/QP solvers are used instead of Lemke's method, then extra techniques should be applied to recover a basis. Appendix \ref{apdx:solver} describes tailored treatments for such cases.}.

Step (3): Verify the uniqueness of mpLCP \eqref{eq:mpLCP_fin}’s solution.
Let $\bar{\Dcal}$ be the indices where $\bar{\v{w}}_{\bar{\Dcal}}(\thetafix) = \bar{\v{z}}_{\bar{\Dcal}}(\thetafix) = \v{0}$. Denote $\bar{\v{M}}_{\bar{\Dcal}, \bar{\Dcal}}$ as a submatrix of $\bar{\v{M}}$ formed by rows and columns both indexed by $\bar{\Dcal}$. Whether $(\bar{\v{w}}(\thetafix),\bar{\v{z}}(\thetafix))$ is the unique solution to mpLCP \eqref{eq:mpLCP_fin} under $\thetafix$ can be identified from the following auxiliary LCP with variable $\v{u}$ \cite[Eq. (2)]{kanekoNumberSolutionsClass1979}
\begin{gather}
    { 
    \begin{gathered}
        \v{u}^\T(\v{L}\v{u} + \v{d}) = 0, \ \v{u} \geq \v{0}, \ \v{L}\v{u} + \v{d} \geq \v{0}, \\
        \v{L} = \begin{bmatrix}
            \bar{\v{M}}_{\bar{\Dcal}, \bar{\Dcal}} & -\bone \\
            \bone^\T & 0
        \end{bmatrix}, \ 
        \v{d} = \begin{bmatrix}
            \v{0} \\
            1
        \end{bmatrix}.
        \label{eq:LCP_aux}
    \end{gathered}
    }
\end{gather}

Let $\v{u}^\star$ be the solution to \eqref{eq:LCP_aux}, $u^\star_{-1}$ be $\v{u}^\star$'s last element, and $z^\star_{0}$ be an ancillary variable introduced by Lemke's method \cite[Eq. (2.3)]{Murty1988-lx}. 
The solution $(\bar{\v{w}}(\thetafix),\bar{\v{z}}(\thetafix))$ is the unique solution to mpLCP \eqref{eq:mpLCP_fin} under $\thetafix$ if one of the two conditions is satisfied:
\begin{enumerate}[leftmargin=*,label=\alph*)]
    \item the auxiliary LCP \eqref{eq:LCP_aux} is infeasible, \ie, $z^\star_{0} > 0$,
    \item the auxiliary LCP \eqref{eq:LCP_aux} is feasible, but $u^\star_{-1} = 0$.
\end{enumerate}
Otherwise, there are infinite many solutions when $u^\star_{-1} > 0$ and $z^\star_{0} = 0$ \cite{kanekoNumberSolutionsClass1979}. As can be seen, when $\bar{\Dcal}$ is empty, the auxiliary LCP \eqref{eq:LCP_aux} is reduced to $u=0$. This indicates $(\bar{\v{w}}(\thetafix),\bar{\v{z}}(\thetafix))$ is a unique solution to mpLCP \eqref{eq:mpLCP_fin}. 

If the solution is nonunique, we will jump to step (3-2): search for all the vertices of mpLCP \eqref{eq:mpLCP_fin}’s solution set under $\thetafix$.
Specifically, if $(\bar{\v{w}}(\thetafix),\bar{\v{z}}(\thetafix))$ is nonunique, then by \cite[Eq. (3)]{kanekoNumberSolutionsClass1979}, the feasible solution set to the elements indexed by ${\bar{\Dcal}}$ in $\v{z}(\thetafix)$ can be written as
\begin{gather}
    { 
    \begin{gathered}
        \Zset_{\bar{\Dcal}}(\thetafix) := \{ 
            \v{z}_{\bar{\Dcal}} \in \Rset^{\vert \bar{\Dcal} \vert}_{+} 
            \ \vert \ 
            \bar{\v{M}}_{\cdot, \bar{\Dcal}} \v{z}_{\bar{\Dcal}} \geq \v{0}, 
            \\
            \ (\bar{\v{M}}_{\bar{\Dcal}, \bar{\Dcal}} + \bar{\v{M}}_{\bar{\Dcal}, \bar{\Dcal}}^\T) \v{z}_{\bar{\Dcal}} = \v{0}
            \}.
            \label{eq:mpLCP_set_zdeg}
    \end{gathered}
    }
\end{gather}

Note the feasible set $\Zset_{\bar{\Dcal}}(\thetafix)$ is polyhedron. 
Denote the vertices of such a polyhedron as $\Vsf[\Zset_{\bar{\Dcal}}, \thetafix]$. Since $\vert \bar{\Dcal} \vert$ is at most $p$, and it is generally much lower than $p$, generation of $\Vsf[\Zset_{\bar{\Dcal}}, \thetafix]$ can be readily done by existing vertex searching techniques. 
Let $\Yset(\thetafix)$ be the solution set of mpLCP \eqref{eq:mpLCP_fin} under $\thetafix$.
Subsequently, the vertices to $\Yset(\thetafix)$, \ie, $\Vsf[ \Yset(\thetafix) ]$ can be obtained as
\begin{gather}
    { 
    \begin{gathered}
        \Vsf[ \Yset(\thetafix) ] := \{
            \v{y} \in \Rset^{2p}_{+} 
            \ \vert \
            \v{y}_{\bar{\Bcal}}(\thetafix) = \bar{\v{M}}_{\cdot, \bar{\Dcal}} \v{z}_{\bar{\Dcal}} + \bar{\v{q}} + \bar{\v{Q}} \thetafix, \\
            \v{y}_{\bar{\Ncal}}(\thetafix) = \v{I}_{\cdot, \bar{\Dcal}} \v{z}_{\bar{\Dcal}},
            \ \forall \v{z}_{\bar{\Dcal}} \in \Vsf[\Zset_{\bar{\Dcal}}, \thetafix] \},
        \label{eq:mpLCP_set_y}
    \end{gathered}
    }
\end{gather}
where $\v{I}$ is an identity matrix. Note if the solution is unique, the complete complementary solution set under $\thetafix$, \ie, $\Yset(\thetafix)$, reduces to a singleton. Consequently, there is only one vertex in $\Vsf[ \Yset(\thetafix) ]$, and it is exactly the solution we have obtained via \eqref{eq:mpLCP.Lemke.ini.sol}. Hence, as indicated in step (3-1), the process in \eqref{eq:mpLCP_set_zdeg}-\eqref{eq:mpLCP_set_y} is no longer needed, and we can save \eqref{eq:mpLCP.Lemke.ini.sol} as $\Vsf[ \Yset(\thetafix) ]$ directly.

{Step (4): Enumerate complementary bases to each vertex of mpLCP’s solution set under the given parameter.}
Recall $\v{y} = \left [ \v{w}^\T, \ \v{z}^\T \right ]^\T$, let $\hat{\Dcal}$ be the indices where $\hat{\v{w}}_{\hat{\Dcal}}(\thetafix) = \hat{\v{z}}_{\hat{\Dcal}}(\thetafix) = \v{0}$ for each vertex $\hat{\v{y}}(\thetafix)$ in $\Vsf[ \Yset(\thetafix) ]$.
By \cite[p64, Def.]{Murty1988-lx}, $\hat{\v{y}}(\thetafix)$ is said to be degenerate if $\hat{\Dcal}$ is non-empty.
Analogous to the Simplex method, where degenerate vertex solution has non-unique bases. 
The complementary basis is also non-unique for the degenerate vertex in LCP. 
We employ a partial enumeration strategy to find all the bases to the mpLCP \eqref{eq:mpLCP_fin} under $\thetafix$. 
Specifically, for the $j^{\textrm{th}}$ vertex $\v{y}_j(\thetafix)$ in $\Vsf[ \Yset(\thetafix) ]$, let a nonoverlapping partition be
\begin{gather} 
    { 
    \begin{gathered}
        \Wcal_j \cup \Zcal_j \cup \Dcal_j = \left \{ 1, 2, \ldots, p \right \},
    \end{gathered}
    }
\end{gather}
where we require
\begin{gather} 
    { 
    \begin{gathered}
        w_k > 0, z_k = 0, \quad \forall \ k \in \Wcal_j, \\
        w_k = 0, z_k > 0, \quad \forall \ k \in \Zcal_j, \\
        w_k = z_k = 0, \quad \forall \ k \in \Dcal_j.
    \end{gathered}
    }
\end{gather}

The indices set $\Wcal_j$, $\Zcal_j$ can help us to identify where we start the enumeration in Fig. \ref{fig:tree}, namely, the index $\{ \Wcal_j, \Zcal_j + p \}$ locates in level $\vert \Wcal_j \vert + \vert \Zcal_j \vert$. 
All possible complementary bases of $\v{y}_j(\thetafix)$ can be express as
\begin{gather}
    { 
    \begin{gathered}
    \Psf[ \v{y}_j(\thetafix) ] := \{ \Wcal_j, \Zcal_j + p, \Scal_j \}, \\
    \Scal_j := \{ k,l \ \vert \ 
        \forall k,l, \ k \in \Dcal_j \land l \in \Dcal_j + p, \\
        k \neq l + p, \ \vert \Scal_j \vert = \vert \Dcal_j \vert \},
    \label{eq:mpLCP.bases.single}
    \end{gathered}
    }    
\end{gather}
which is $2^{\vert \Dcal_j \vert}$ candidates in total. All feasible complementary bases of $\v{y}_j(\thetafix)$ can be express as
\begin{gather}
    \Bsf[ \v{y}_j(\thetafix) ] := \{ \Bcal_j \ \vert \ \rank \v{Y}_{\cdot, \Bcal_j} = p, \ \forall \Bcal_j \in \Psf[ \v{y}_j(\thetafix) ] \}.
    \label{eq:mpLCP.rank}
\end{gather}

The rank check is to ensure all basis matrix $\bar{\v{B}}$ must be invertible to generate the coefficients in \eqref{eq:mpLCP.Lemke.ini.coef} when $\bar{\v{B}} = \v{Y}_{\cdot, \Bcal_j}$. 
Denote $\Bsf[ \Yset(\thetafix) ]$ as the set of complementary basis which has been enumerated from each vertex in $\Vsf[ \Yset(\thetafix) ]$, we have
\begin{gather}
    \Bsf[ \Yset(\thetafix) ] := \{ \Bsf[ \v{y}(\thetafix) ] \ \vert \ \forall \v{y}(\thetafix) \in \Vsf[ \Yset(\thetafix) ] \}.
    \label{eq:mpLCP.bases.all}
\end{gather}

\begin{figure}[htbp]
    \centering
    \vspace{-10pt}
    \begin{tikzpicture}
        \tikzset{
        font=\scriptsize,
        level 1/.style={level distance=15mm, sibling distance=0mm},
        level 2/.style={level distance=25mm, sibling distance=0mm}, 
        level 3/.style={level distance=30mm, sibling distance=-1mm},               
        grow'=right}                
        \Tree [.Level\ 0 [.Level\ 1 [.Level\ 2 [.Level\ 3 ] ] ] ]            
    \end{tikzpicture}   
    \vspace{4pt}

    \begin{tikzpicture}
        \tikzset{
        font=\scriptsize,
        level 1/.style={level distance=20mm, sibling distance=0mm},
        level 2/.style={level distance=25mm, sibling distance=0mm}, 
        level 3/.style={level distance=30mm, sibling distance=-1mm},               
        grow'=right}        
        \Tree [.{} 
        [.\{$1$\}
        [.\{$1,2$\} 
        [.\{$1,2,3$\} ] [.\{$1,2,6$\} ] ] 
        [.{\color{red}\{$1,5$\}} 
        [.{\color{red}\{$1,3,5$\}} ] 
        [.{\color{red}\{$1,5,6$\}} ] ] ] 
        [.\{$4$\} 
        [.\{$2,4$\} 
        [.\{$2,3,4$\} ] [.\{$2,4,6$\} ] ] 
        [.\{$4,5$\} 
        [.\{$2,3,4$\} ] [.\{$2,4,6$\} ] ] ]
        ]              
    \end{tikzpicture}
    \vspace{-7pt}
    \caption{Generation and selection of basis candidates when $p=3$, $\Wcal_j = 1$, $\Zcal_j = 2$, $\Dcal_j = 3$.}
    \label{fig:tree.illustration}
\end{figure}
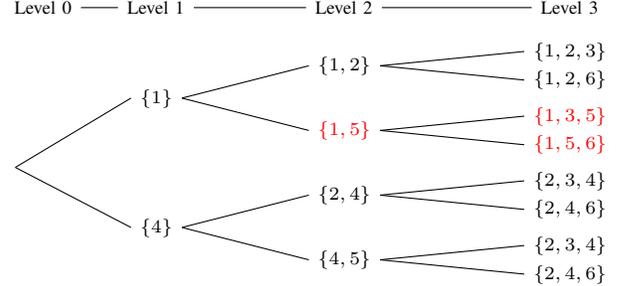

For illustrative purpose, let $p=3$, the index partitions to the $j^{\textrm{th}}$ vertex in $\Vsf[ \Yset(\thetafix) ]$ are $\Wcal_j = 1$, $\Zcal_j = 2$, and $\Dcal_j = 3$. 
Then, as can be clearly seen in Fig. \ref{fig:tree.illustration}, the candidate bases to $j^{\textrm{th}}$ vertex are $\{1, 3, 5\}$ and $\{1, 5, 6\}$. They are the supersets to the index $\{ \Wcal_j, \Zcal_j + p \} = \{ 1, 5 \}$ which lies in the level $\vert \Wcal_j \vert + \vert \Zcal_j \vert = 2$. We can see that up to six out of eight candidates have been pruned instantly in level $p=3$. 
Candidates of $\Bcal_j$, \ie, $\{1, 3, 5\}$ and $\{1, 5, 6\}$, can be further pruned using \eqref{eq:mpLCP.rank}.
We acknowledge calculating rank may be costly when the matrix is large. Nevertheless, the modern computer can conduct such computations quite efficiently. The accelerated rank calculation, such as involving sparse techniques, is beyond the scope of this paper. We leave interested readers for faster practical implementation. 

Step (5): Obtain all critical regions containing $\thetafix$ using all complementary bases.
Let $\Bcal_k$ be the $k^{\textrm{th}}$ basis in $\Bsf[ \Yset(\thetafix) ]$, by \cite[Eq. (4)-(5)]{hercegEnumerationbasedApproachSolving2015a}, the parametric form to the basic and nonbasic variables under $\Bcal_k$ admits the following form
\begin{gather} 
    { 
    \begin{gathered}
    \v{y}_{\Bcal_k}(\v{\theta}; \Bcal_k) := \bar{\v{q}} + \bar{\v{Q}} \v{\theta}, 
    \quad
    \v{y}_{\Ncal}(\v{\theta}; \Bcal_k) := \v{0},    
    \label{eq:mpLCP.Lemke.ini.map}
    \end{gathered}
    }
\end{gather}
where the calcution of $\bar{\v{q}}$, $\bar{\v{Q}}$ can refer to \eqref{eq:mpLCP.Lemke.ini.coef} by letting $\bar{\Bcal} = \Bcal_k$. 
The set of parameters for which $\v{y}_{\Bcal_k}(\v{\theta}; \Bcal_k) \geq \v{0}$ then forms the critical region $\CR\[\Bcal_k\]$, \ie,
\begin{gather} 
    { 
    \begin{gathered}
        \CR\[\Bcal_k\]
        := \{ \v{\theta} \ \vert \ \v{y}_{\Bcal_k}(\v{\theta}; \Bcal_k) \geq \v{0} \}, 
        \label{eq:mpLCP.Lemke.map.CR}
    \end{gathered}
    }
\end{gather}
and the value function $\JJ(\v{\theta}, \Bcal_k)$ defined over $\CR[\Bcal_k]$ is
\begin{gather} 
    { 
    \begin{gathered}
    \JJ(\v{\theta}; \Bcal_k)
        := \frac{1}{2} \v{x}^\T(\v{\theta}; \Bcal_k) \v{H} \v{x}(\v{\theta}; \Bcal_k) + \v{f}^\T \v{x}(\v{\theta}; \Bcal_k),
        \label{eq:mpLCP.Lemke.map.VF}
    \end{gathered}
    }
\end{gather}
where $\v{x}(\v{\theta}; \Bcal_k)$ is obtained by taking the corresponding rows in $\v{y}(\v{\theta}; \Bcal_k)$. 
For mpLCP \eqref{eq:mpLCP_fin} under $\thetafix$, $\Bsf[ \Yset(\thetafix) ]$ describes all the bases to the basic solutions in $\Vsf[ \Yset(\thetafix) ]$. And there is an one-to-one correspondence between each basis $\Bcal_k$ in $\Bsf[ \Yset(\thetafix) ]$ to the critical region $\CR\[\Bcal_k\]$. Consequently, the union of $\CR$ generated from $\Bsf[ \Yset(\thetafix) ]$ fully partitioned the parameter space in a neighborhood of $\thetafix$.

The complete process to search for all critical regions containing $\thetafix$ is summarized in Algo. \ref{alg:LC-Pivoting}. 
The algorithm first reformulates mpLP/QP into mpLCP in step \ref{alg:LC-Pivoting.ini}. 
For mpLCP \eqref{eq:mpLCP_fin} under $\thetafix$, it is simplified into an LCP and can be solved by Lemke's method. 
The basis $\bar{\Bcal}$, basic complementary solutions $\bar{\v{y}}(\thetafix)$ and coefficients $\bar{\v{M}}$, $\bar{\v{q}}$, $\bar{\v{Q}}$ are obtained in steps \ref{alg:LC-Pivoting.Lemke1}-\ref{alg:LC-Pivoting.Lemke3}.  
If the zero indices $\bar{\Dcal}$ is empty from step \ref{alg:LC-Pivoting.Lemke4}, then unique vertex and basis are returned in steps \ref{alg:LC-Pivoting.nodeg1}-\ref{alg:LC-Pivoting.nodeg2}. 
Otherwise, we will verify the uniqueness of mpLCP \eqref{eq:mpLCP_fin} 's solution under $\thetafix$ in step \ref{alg:LC-Pivoting.deg.lcp}. And the vertices to the solution set, \ie, $\Vsf[ \Yset(\thetafix) ]$, is obtained in steps \ref{alg:LC-Pivoting.deg.unique1}-\ref{alg:LC-Pivoting.deg.nonunique2}. 
In step \ref{alg:LC-Pivoting.deg.basis}, the bases to each basic complementary solution in $\Vsf[ \Yset(\thetafix) ]$ is enumerated, and the basis set is denoted as $\Bsf[ \Yset(\thetafix) ]$. 
Finally, in step \ref{alg:LC-Pivoting.map}, all critical regions containing $\thetafix$ along with the corresponding value functions defined over them are denoted as $\CR[\thetafix]$, $\JJ(\v{\theta}; \thetafix)$. And they are generated by \eqref{eq:mpLCP.Lemke.map.CR}-\eqref{eq:mpLCP.Lemke.map.VF} from each basis $\Bcal_k$ in $\Bsf[ \Yset(\thetafix) ]$.

\begin{algorithm}[!htbp]
    \caption{Search for all $\CR$ containing $\thetafix$.}
    \label{alg:LC-Pivoting}

    Set $\thetafix$, $\Bsf[ \Yset(\thetafix) ] \gets \emptyset$, $\Vsf[ \Yset(\thetafix) ] \gets \emptyset$.
    \label{alg:LC-Pivoting.ini}

    $\mpLCP (\v{M}, \v{q}, \v{Q}) \gets$ reformulation of mpLP/QP \eqref{eq:mpLCP_ini}
    \label{alg:LC-Pivoting.mpLCP}

    $\bar{\Bcal} \gets$ basis from Lemke's complementary pivot rule
    \label{alg:LC-Pivoting.Lemke1}

    $\bar{\v{y}}(\thetafix) \gets$ complementary solution under $\bar{\Bcal}$
    \label{alg:LC-Pivoting.Lemke2}

    $\bar{\v{M}}$, $\bar{\v{q}}$, $\bar{\v{Q}} \gets$ final coefficients in Lemke's method
    \label{alg:LC-Pivoting.Lemke3}
    
    $\bar{\Dcal} \gets$ indices with $\bar{\v{w}}_{\bar{\Dcal}}(\thetafix) = \bar{\v{z}}_{\bar{\Dcal}}(\thetafix) = \v{0}$
    \label{alg:LC-Pivoting.Lemke4}

    \eIf(){$\bar{\Dcal} = \emptyset$}{
        \label{alg:LC-Pivoting.nodeg1}
        $\Vsf[ \Yset(\thetafix) ] \gets \bar{\v{y}}(\thetafix)$ and $\Bsf[ \Yset(\thetafix) ] \gets \bar{\Bcal}$
        \label{alg:LC-Pivoting.nodeg2}
    }(){
        $z^\star_{0}, u^\star_{-1}$ solution to auxiliary LCP \eqref{eq:LCP_aux}
        \label{alg:LC-Pivoting.deg.lcp}

        \eIf(){$z^\star_{0} > 0$ and $u^\star_{-1} = 0$}{
            \label{alg:LC-Pivoting.deg.unique1}
            $\Vsf[ \Yset(\thetafix) ] \gets \bar{\v{y}}(\thetafix)$
            \label{alg:LC-Pivoting.deg.unique2} 
        }(){
            \label{alg:LC-Pivoting.deg.nonunique1}
            $\Vsf[ \Yset(\thetafix) ] \gets$ vertex search for \eqref{eq:mpLCP_set_zdeg}-\eqref{eq:mpLCP_set_y}
            \label{alg:LC-Pivoting.deg.nonunique2}
        }
        $\Bsf[ \Yset(\thetafix) ] \gets$ bases enumeration via \eqref{eq:mpLCP.bases.all}
        \label{alg:LC-Pivoting.deg.basis}
    }
    \Return $\CR[\thetafix]$, $\JJ(\v{\theta}; \thetafix)  \gets$ all $\CR$s and value functions generated from each $\Bcal_k \in \Bsf[ \Yset(\thetafix) ]$ using \eqref{eq:mpLCP.Lemke.map.CR}-\eqref{eq:mpLCP.Lemke.map.VF}
    \label{alg:LC-Pivoting.map}
\end{algorithm}

\section{An Improved Critical Region Exploration}
\label{sec:improved_CRE}

\subsection{Distributed problem setup}

To show the effectiveness and practical application of the proposed degeneracy handling method, we integrate the Algo. \ref{alg:CRE_ARS} into CRE method \cite{guoRobustTieLineScheduling2018}. Arm with two other nontrivial modifications to CRE, we apply the improved CRE to solve the tie-line scheduling problem.

We begin by considering a general model of distributed convex optimization for multi-agent systems\footnote{
    We use the term [distributed] to entail coordinated/layered and fully decentralized structures for the multi-agent system. The former has a central coordinator, whereas the latter does not.
    }. 
The agents are with local decision variables $\v{x}_i$, $i = 1, \ldots, N$ and coupled via variable $\v{\theta}$. 
The optimization model can be formulated as a linear constrained quadratic programming problem.
\begin{subequations}
    \label{eq:probP}
    \begin{alignat}{2}
    &&  
    \underset{\v{x}_i \in \Rset^{n_i}, \v{\theta} \in \Rset^d}{\text{minimize}} & 
    \quad \sum_{i=1}^{N} \( \frac{1}{2} \v{x}_i^\T \v{H}_i \v{x}_i + \v{f}_i^\T \v{x}_i \), 
    \label{eq:probP.obj} \\ 
    && \text{subject to} &
    \quad \v{A}_i \v{x}_i \leq \v{b}_i + \v{C}_{i} \v{\theta},  \ \forall \ i = 1 , \dots, N,
    \label{eq:probP.local} \\
    &&&	
    \quad \v{\Theta}_i := \{ \v{D}_{i} \v{\theta} \leq \v{r}_i \}, \ \forall \ i = 1 , \dots, N.
    \label{eq:probP.couple}
    \end{alignat}
\end{subequations}

The objective \eqref{eq:probP.obj} is a convex function with $\v{H}_i \succeq 0$. 
Variables $\v{x}_i$ are restricted in constraint \eqref{eq:probP.local}. 
Constraint \eqref{eq:probP.couple} bounds coupling variables $\v{\theta}$. 
For notational convenience, we replace all the equality constraints with inequalities in \eqref{eq:probP.local}-\eqref{eq:probP.couple}. 
By projecting $\v{x}_i$ onto the $\v{\theta}$ space, problem \eqref{eq:probP} can be equivalently rewritten as
\begin{subequations}
    \label{eq:probP_re}
    \begin{alignat}{2}
    &&  
    \underset{\v{\theta} \in \Rset^d}{\text{minimize}} & 
    \quad \JJ(\v{\theta}) := \textstyle \sum_{i=1}^N \JJ_i(\v{\theta}),
    \label{eq:probP_re.obj} \\ 
    && \text{subject to} &
    \quad \v{\theta} \in \v{\Theta}^\star := \cap_{i=1}^N \v{\Theta}^\star_i,
    \label{eq:probP_re.local} \\
    &&&	
    \quad \v{\theta} \in \v{\Theta} := \cap_{i=1}^N \v{\Theta}_i,
    \label{eq:probP_re.couple}
    \end{alignat}
\end{subequations}
where for each $\v{\Theta}^\star_i$, it is defined as
\begin{alignat}{2}
    & \v{\Theta}_i^\star := \{ \v{\theta} \in \Rset^d \ \vert \ \exists \ \v{x}_i: \v{A}_i \v{x}_i \leq \v{b}_i + \v{C}_i \v{\theta} \}, 
    \label{eq:fea_para_sp}
\end{alignat}
and the $\JJ_i(\v{\theta})$ is given as
\begin{subequations}
    \label{eq:mpP}
    \begin{alignat}{3}
    \JJ_i(\v{\theta}) := \ 
    & \underset{\v{x}_i \in \Rset^{n_i}}{\text{minimize}} 
    && \ \frac{1}{2} \v{x}_i^\T \v{H}_i \v{x}_i + \v{f}_i^\T \v{x}_i, 
    && 
    \label{eq:mpP.obj} \\ 
    & \text{subject to}
    && \ \v{A}_i \v{x}_i \leq \v{b}_i + \v{C}_i \v{\theta},
    &&
    \label{eq:mpP.cons}
    \end{alignat}
\end{subequations}
which is exactly the general mpQP formulation \eqref{eq:mpSemiQP} when $\v{\theta}$ is treated as parameters.
The property of the value function $\JJ_i(\v{\theta})$ is summarized in the following lemma.

\begin{lemma}
    (\cf \cite[Th. 6.5, 6.7]{borrelli_bemporad_morari_2017}):
    \label{lemma.mpP}
    Consider the general convex mpQP \eqref{eq:mpP}, the set of feasible parameters $\v{\Theta}^\star_i$ is polyhedral
    and $\JJ_i(\v{\theta})$ is convex and piecewise linear/quadratic over $\v{\Theta}^\star_i$. 
\end{lemma}

Since $\JJ_i(\v{\theta})$ is a piecewise function, each segment of $\JJ_i(\v{\theta})$ corresponds to a subset of feasible space $\v{\Theta}^\star_i$. In particular, such a subset corresponds to a critical region we have already defined in \eqref{eq:KKT_CR}. The following two corollaries naturally follow.

\begin{corollary}
    \label{corollary:multi_VF}
    The overall value function $\JJ(\v{\theta})$ is convex and piecewise linear/quadratic. The segments of $\JJ(\v{\theta})$ are defined over a group of polyhedral $\CR$s. Each $\CR = \v{\Theta} \cap_{i=1}^N \CR_i$ is an intersection of local critical regions and initial feasible space $\v{\Theta}$.
\end{corollary}

\begin{corollary}
    \label{corollary:mpLCP}
    If the general convex mpQP \eqref{eq:mpP} is reformulated into mpLCP \eqref{eq:mpLCP_fin}, lemma \ref{lemma.mpP} and corollary \ref{corollary:multi_VF} still hold for the critical region and value function obtained via \eqref{eq:mpLCP.Lemke.map.CR}-\eqref{eq:mpLCP.Lemke.map.VF}.
\end{corollary}

\subsection{General critical region exploration process}

\begin{figure}[htbp]
    \centering
    \vspace{-12pt}
    \includegraphics[width=0.4\textwidth]{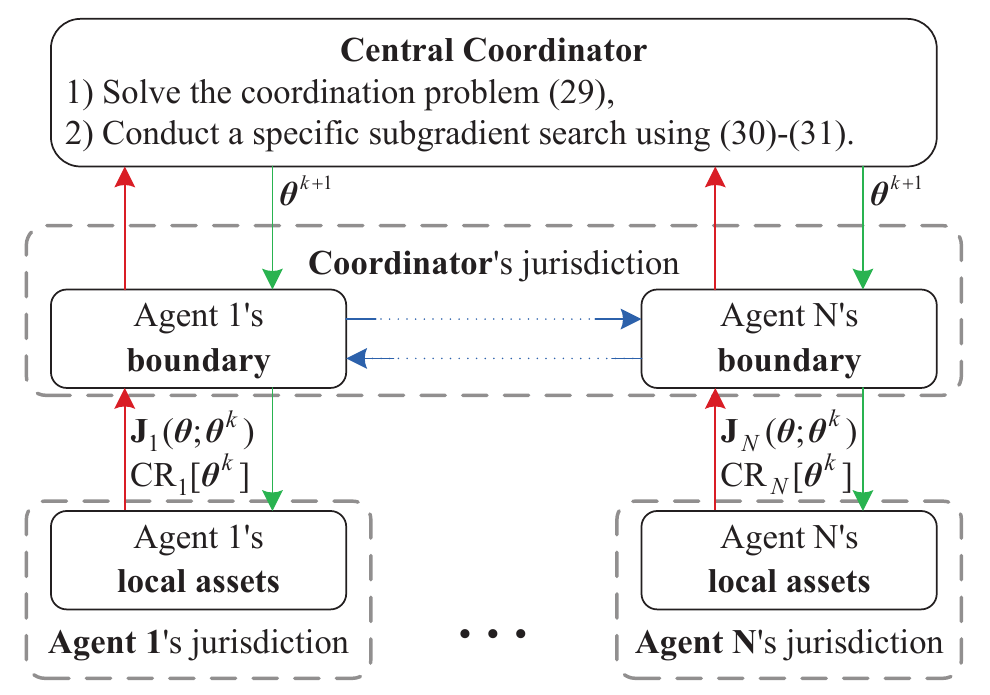}    
    \vspace{-7pt} 
    \caption{The architecture of CRE. Red/green arrows indicate upward/downward communication links and blue arrows indicate physical connections.}
    \label{fig:arch_cre}
\end{figure}

The CRE adopts a two-layer structure. As shown in Fig. \ref{fig:arch_cre}, a central coordinator resides on the upper level, which optimizes all agents' boundary states $\v{\theta}$. Similar to the existing primal decomposition-based approach, where the local variables $\v{x}_i, i = 1, \ldots, N$ and coupling variable $\v{\theta}$ are iteratively updated. The advantage of the CRE coordination is that it finds the value function and $\CR$ \wrt $\v{\theta}$ to accelerate the $\v{\theta}$'s update. 

Under Lemma \ref{lemma.mpP} and the structure of Fig. \ref{fig:arch_cre}, CRE recursively solves problem \eqref{eq:probP} by applying the following two steps. 
\begin{enumerate}[leftmargin=*]
    \item \emph{Local evaluation:} Given $\thetak$, each agent solves problem \eqref{eq:mpP} to get $\JJ_i(\v{\theta}; \thetak)$, $\CR_i[\thetak]$ from \eqref{eq:KKT_CR}-\eqref{eq:KKT_VF}. 
    \item \emph{Coordination update:} First, solves the problem below.
\end{enumerate}
\begin{subequations}
    \label{eq:cre.coordination}
    \begin{alignat}{2}
    &  
    \textstyle \text{minimize}_{\v{\theta} \in \v{\Theta}}
    && 
    \ \JJ(\v{\theta}; \thetak) := \textstyle \sum_{i=1}^N \JJ_i(\v{\theta}; \thetak), \\
    & \text{subject to} &&
    \ \CR\[\thetak\] := \cap_{i=1}^N \CR_i\[\thetak\],
    \end{alignat}
\end{subequations}
to obtain the optimal solution pair $(\thetaopt, \JJ(\thetaopt))$ in the current critical region $\CR\[\thetak\]$. Then $\thetaopt$ is updated by a projected subgradient search 
\begin{alignat}{2}
    & \v{\theta}^{k+1} := \thetaopt - \varepsilon \v{v}^\star,
    \label{eq:CRE.search}
\end{alignat}
where $\varepsilon$ is a small stepsize, $\v{v}^\star$ is the optimal $\v{v}$ from
\begin{subequations}
    \label{eq:CRE.LSQ}
    \begin{alignat}{2}
    &
    \textstyle \text{minimize}_{\v{v}, \v{\eta}, \v{\zeta}}
    && 
    \ \|\v{v}\|^2, \\ 
    & \text{subject to} &&
    \ \v{v} = (\partial \JJ^\star) \v{\eta} + (\Nsf\[\thetaopt; \v{\Theta}\]) \v{\zeta}, \\
    &&& \ \bone^\T \v{\eta} = 1, \ \v{\eta} \geq \v{0}, \ \v{\zeta} \geq \v{0}.
    \end{alignat}
\end{subequations}
Here, an all-one vector is denoted by $\bone$. 
Notations $\v{\eta}$ and $\v{\zeta}$ are weight variables for the subdifferential set $\partial \JJ^\star$ and normal cone $\Nsf\[\thetaopt; \v{\Theta}\]$. 
The true subdifferential set $\partial \JJ\[\thetaopt\]$ at $\thetaopt$ is unknown, which is replaced by the subdifferential set obtained from past iterations, \ie, $\partial \JJ^\star = \cup_{t=1}^k \partial \JJ \[\thetaopt; \CR\[\v{\theta}^t\] \]$. 
We remark $\partial \JJ \[\thetaopt; \CR\[\v{\theta}^t\] \] = \emptyset$ if $\thetaopt \notin \CR\[\v{\theta}^t\]$. 
Hence, such $\CR\[\v{\theta}^t\]$ will be discarded during iterations. 
CRE converges to the exact optima of \eqref{eq:probP} after finite iterations when $\|\v{v}\| = 0$, detailed proofs could refer to \cite{guoRobustTieLineScheduling2018}. 
The optimal solution to \eqref{eq:probP} evaluated at the optimal boundary state $\thetaopt$ is $\JJ(\thetaopt)$.

\subsection{An improved scheme}
\label{sec:CRE_AutoR}

By leveraging mpLCP formulation to deal with degeneracy, the CRE can proceed under complex degenerate situations. An improved CRE process is detailed in Algo. \ref{alg:CRE_ARS}.
The algorithm is initialized with a start point $\v{\theta}^{0} \in \v{\theta}$ and a default stepsize is set as $\varepsilon = 10^{-2}$. 
We highlight three improvements for the CRE process in Algo. \ref{alg:CRE_ARS}.
First, we incorporate Algo. \ref{alg:LC-Pivoting} into CRE. 
In steps \ref{alg:CRE_ARS.LC1}-\ref{alg:CRE_ARS.LC3}, we convert the local problem \eqref{eq:mpP} into mpLCP and solve it under $\v{\theta}^{k}$. 
Under such a scheme, CRE does not rely on equations \eqref{eq:KKT_CR}-\eqref{eq:KKT_VF} to generate critical regions and value functions. Instead, equations \eqref{eq:mpLCP.Lemke.map.CR}-\eqref{eq:mpLCP.Lemke.map.VF} are adopted for any $\v{\theta}^{k+1}$.
In steps \ref{alg:CRE_ARS.LC4}-\ref{alg:CRE_ARS.LC5}, multiple $\CR$ might be generated when there is degeneracy.
Hence, problem \eqref{alg:CRE_ARS.co.opt} might also be solved multiple times in step \ref{alg:CRE_ARS.co.opt} under each CR. The optimal solution to the coordination problem \eqref{alg:CRE_ARS.co.opt} is denoted by $(\thetatmp$, $\JJ(\thetatmp))$. 
In this way, CRE can proceed when degenerate situations are encountered.

\begin{algorithm}[!htbp]
    \caption{An improved CRE process.}
    \label{alg:CRE_ARS}

    Set 
    $\v{\theta}^\star \in \v{\Theta}$,
    $\JJ(\v{\theta}^\star) \gets 10^{8}$,
    $k \gets 0$,
    $\v{\theta}^{k} \gets \v{\theta}^\star$,
    $\varepsilon = 10^{-2}$,
    $\epsilon = 10^{-4}$, 
    $\alpha$, $\beta$ are pre-defined coefficients.

    \For{$k=0, 1, 2, \ldots$}{

        \For{$i=1, \ldots, N$}{
            \label{alg:CRE_ARS.LC1}

            $\bar{\v{y}}_i(\v{\theta}^{k})$, $z_{0, i}^\star \gets$ solution to \eqref{eq:mpLCP_fin} under $\v{\theta}^{k}$
            \label{alg:CRE_ARS.LC2}

            \lIf(){$\vert z_{0, i}^\star \vert > 0$}{
                $\v{\theta}^{k}$ is infeasible 
                \label{alg:CRE_ARS.LC3}
                }
        }

        \eIf(){$\v{\theta}^{k}$ is infeasible}{
            \label{alg:CRE_ARS.FC1}
            \For{$i=1, \ldots, N$}{
                $\FC_i[\v{\theta}^{k}] \gets$ cutting plane from \eqref{eq:mpP_fea}-\eqref{eq:mpP_fea.FC}
                \label{alg:CRE_ARS.FC2}
            }
            $\v{\Theta} \gets \v{\Theta} \cap_{i=1}^N \FC_i[\v{\theta}^{k}]$,
            $\v{\theta}^{k+1} \gets \thetaproj$ from \eqref{eq:mpP_fea.proj}
            \label{alg:CRE_ARS.FC3}
        }{
            \For{$i=1, \ldots, N$}{
                \label{alg:CRE_ARS.LC4}
                $\JJ_i(\v{\theta}; \v{\theta}^{k})$, $\CR_i\[\v{\theta}^{k}\] \gets$ value functions and critical regions by conducting Algo. \ref{alg:LC-Pivoting}         
                \label{alg:CRE_ARS.LC5}
            }
        $\thetatmp$, $\JJ(\thetatmp) \gets$ optimal solution of \eqref{eq:cre.coordination}
        \label{alg:CRE_ARS.co.opt} 

        \uIf(){$\JJ(\v{\theta}^\star) - \JJ(\thetatmp) \geq \epsilon$}{
            \label{alg:CRE_ARS.co.jdg1}
            $\varepsilon_k \gets \varepsilon$, 
            $\JJ(\v{\theta}^\star) \gets \JJ(\thetatmp)$,
            $\v{\theta}^\star \gets \thetatmp$, 
            $\partial \JJ^\star \gets \partial \JJ\[\v{\theta}^\star, \CR\[\v{\theta}^{k}\] \]$
            \label{alg:CRE_ARS.co.jdg2}
        }
        \uElseIf{$-\epsilon \leq \JJ(\v{\theta}^\star) - \JJ(\thetatmp) < \epsilon$}{
            \label{alg:CRE_ARS.co.jdg3}
            $\varepsilon_k \gets \min(\alpha \varepsilon_{k-1}, \varepsilon)$, 
            $\partial \JJ^\star \gets \partial \JJ^\star \cup \partial \JJ\[\v{\theta}^\star, \CR\[\v{\theta}^{k}\] \]$
            \label{alg:CRE_ARS.co.jdg4}
        }
        \Else{
            \label{alg:CRE_ARS.co.jdg5}
            $\varepsilon_k \gets \max(\beta \varepsilon_{k-1}, 10^{-5})$      
            \label{alg:CRE_ARS.co.jdg6}
        }

        $\Nsf\[\v{\theta}^\star; \v{\Theta}\] \gets$ updated normal cone
        \label{alg:CRE_ARS.co.expl0}

        $\v{v}^\star \gets$ projected subgradient of \eqref{eq:CRE.LSQ}
        \label{alg:CRE_ARS.co.expl1}         
        
        \eIf(){$\lVert \v{v}^\star \rVert = 0$}{
            \label{alg:CRE_ARS.co.return1}
            \textbf{break}
            \label{alg:CRE_ARS.co.return2}
        }{
            $\v{\theta}^{k+1} \gets \v{\theta}^\star - \varepsilon_k \v{v}^\star$
            \label{alg:CRE_ARS.co.expl2}
        }
        }
    }
    \Return $\v{\theta}^\star$, $\JJ(\v{\theta}^\star)$
    \label{alg:CRE_ARS.co.return3}
\end{algorithm}

Second, we add a cutting plane scheme to ensure $\v{\theta}^{k+1}$ is feasible to local mpLP/QP \eqref{eq:mpP} (or equivalently, local mpLCP in \eqref{eq:mpLCP_fin}) after finite iterations. 
Recall $\v{\Theta}$ is the intersection of all coupling constraints \eqref{eq:probP.couple}. 
By equation \eqref{eq:fea_para_sp}, $\v{\Theta}^\star$ is the region of $\v{\theta}$ where the local problem \eqref{eq:mpP} has feasible solutions.
Due to privacy and computational efficiency concerns, the coordinator does not know the true $\v{\Theta}^\star$. 
Hence, the following feasibility problem \cite[Sec 5.1b]{birgeIntroductionStochasticProgramming2011} derived from problem \eqref{eq:mpP} is solved when $\v{\theta}^{k+1} \notin \v{\Theta}^\star$ after the subgradient update.
\begin{subequations}
    \label{eq:mpP_fea}
    \begin{alignat}{3}
    & \underset{\v{x}_i, \v{s}_{i}}{\text{minimize}}
    && \ \bone^\T \v{s}_{i}, 
    && 
    \\ 
    & \text{subject to}
    && \ \v{A}_i \v{x}_i - \v{s}_{i} \leq \v{b}_i + \v{C}_{i} \v{\theta}^{k+1},
    && : \v{\lambda}_i \in \Rset^{m_i} 
    \label{eq:mpP_fea.cons} 
    \\
    &
    && \ \v{s}_{i} \geq \v{0}.
    && 
    \end{alignat}
\end{subequations}

Denote the optimal solution to \eqref{eq:mpP_fea} as $\v{x}^\star_i$ and the optimal multipliers to \eqref{eq:mpP_fea.cons} as $\v{\lambda}^\star_i$. Each agent's feasibility cut of $\v{\theta}$ at $\v{\theta}^{k+1}$ is generated by
\begin{gather}
    \FC_i[\v{\theta}^{k+1}] := \{ \v{\theta} \ \vert \ \v{\lambda}^{\star, \T}_i(\v{A}_i \v{x}_i^\star - \v{b}_i) \leq \v{\lambda}^{\star, \T}_i \v{C}_{i} \v{\theta} \}.
    \label{eq:mpP_fea.FC}
\end{gather}

The coordinator can now update $\v{\Theta}$ by $\tilde{\v{\Theta}} := \v{\Theta} \cap_{i=1}^N \FC_i[\v{\theta}^{k+1}]$. 
And a new parameter can be obtained via projecting $\v{\theta}^{k+1}$ onto the new parameter space $\tilde{\v{\Theta}}$, \ie
\begin{alignat}{2}
    \thetaproj := 
        \{  
        \textstyle \text{argmin}_{\v{\theta}} \ 
        \lVert \v{\theta} - \v{\theta}^{k+1} \rVert 
        \ \vert \  
        \v{\theta} \in \tilde{\v{\Theta}} \}.
    \label{eq:mpP_fea.proj}
\end{alignat}

The feasibility cuts are generated in steps \ref{alg:CRE_ARS.FC1}-\ref{alg:CRE_ARS.FC3}.
We also leverage the ancillary variables to Lemke's method for certification of the infeasibility \cite[Eq. (2.3)]{Murty1988-lx}.
When the new $\v{\theta}^{k}$ becomes feasible, each agent leverages Algo. \ref{alg:LC-Pivoting} to obtain value functions $\JJ_i(\v{\theta}; \v{\theta}^{k})$ and critical regions $\CR_i\[\v{\theta}^{k}\]$ in steps \ref{alg:CRE_ARS.LC4}-\ref{alg:CRE_ARS.LC5}.

Finally, adaptive adjustments of stepsize in \eqref{eq:CRE.search} is designed to search for adjacent $\CR$s.
The requirement of such $\varepsilon$ to explore parameter space is that $\varepsilon$ should be large enough to step into an adjacent $\CR$ but not cross over it. 
Hence, selecting $\varepsilon$ is tricky, as some $\CR$ may be small or flat. 
An inappropriate stepsize may cause cycling of CRE when some important $\CR$ have been missed.
In Algo. \ref{alg:CRE_ARS}, the coordinator adjusts stepsizes in steps \ref{alg:CRE_ARS.co.jdg1}-\ref{alg:CRE_ARS.co.jdg6}, along with the updates to the current optimal solutions. We distinguish three cases:
\begin{enumerate}[leftmargin=*]
    \item If the new solution is better (steps \ref{alg:CRE_ARS.co.jdg1}-\ref{alg:CRE_ARS.co.jdg2}), we will update optimal solution ($\v{\theta}^\star$, $\JJ(\v{\theta}^\star)$) just like the classic CRE. 
    The differences are 
    (a) $\varepsilon_k$ is reset to the initial values, 
    (b) the approximate subdifferential $\partial \JJ^\star$ is initialized by subdifferential $\partial \JJ\[\v{\theta}^\star, \CR\[\v{\theta}^{k}\] \]$ as multiple subgradients can be calculated if there are degenerate situations,
    \item If the new solution is the same (steps \ref{alg:CRE_ARS.co.jdg3}-\ref{alg:CRE_ARS.co.jdg4}), indicating we explored at least a new $\CR$ that is not better than the current. Then we will adjust the stepsize by a pre-defined coefficient and append the newly explored subdifferential,
    \item If the new solution is worse than the current (steps \ref{alg:CRE_ARS.co.jdg5}-\ref{alg:CRE_ARS.co.jdg6}), we may step over some vital $\CR$s. Hence, we will only decrease stepsize and make no change to the current $\v{\theta}^\star$.
\end{enumerate}

After the above process, we check if there are updates in the normal cone $\Nsf\[\v{\theta}^\star; \v{\Theta}\]$. 
And then, we apply a projected subgradient search in steps \ref{alg:CRE_ARS.co.expl1} and \ref{alg:CRE_ARS.co.expl2}. 
As shown in steps \ref{alg:CRE_ARS.co.return1}-\ref{alg:CRE_ARS.co.return2}, the proposed scheme also terminates to an exact solution when the gradient norm $\lVert \v{v}^\star \rVert$ is less than a threshold. 
And the optimal solutions are returned in step \ref{alg:CRE_ARS.co.return3}.
The above modifications enable CRE with better practical performance and convergence.

\subsection{Application to multi-area tie-line scheduling}

\begin{figure}[htbp]
    \centering
    \vspace{-15pt}
    \includegraphics[width=0.4\textwidth]{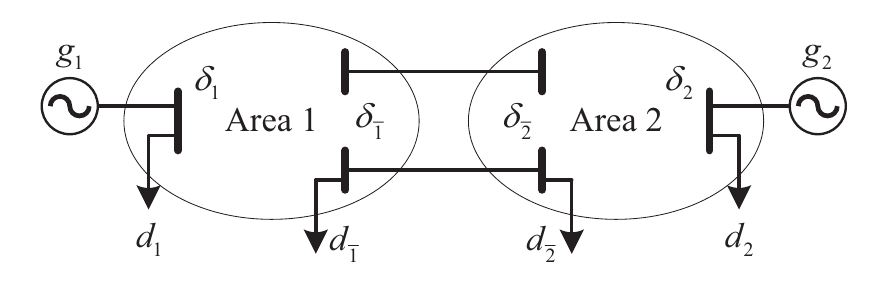}    
	\vspace{-15pt}
    \caption{An illustration for the multi-area power system.}
    \label{fig:maopf}
\end{figure}

Problem \eqref{eq:probP} can adapt to many formulations. 
This paper considers an application scenario of multi-area tie-line scheduling. 
To formulate the multi-area tie-line scheduling problem, we adopt a DC power flow model \cite[Eq. (6.40)-(6.41)]{Wood2013-eq} for each SO's transmission network, which has been proved to be a good linear approximation. 
Let the DC approximated nodal and branch admittance matrix be denoted as $\v{B}$, $\v{H}$. 
The subscripts of the matrices reflect the subdivision of buses and transmission lines based on geographic features. 
An $N$-area tie-line scheduling problem can be formulated as
\begin{subequations}
	\label{eq:maopf_theta}
	\begin{alignat}{2}
	\underset{\v{g}_i, \v{\delta}_i, \v{\delta}_{\bar{i}}}{\text{minimize}} & 
	\quad \sum_{i=1}^{N} \( \frac{1}{2} \v{g}_i^\T \v{Q}_i \v{g}_i + \v{c}_i^\T \v{g}_i \), 
	\label{eq:maopf_theta.obj} \\ 
	\text{subject to} &
	\quad \v{B}_{i,i} \v{\delta}_i + \v{B}_{i,\bar{i}} \v{\delta}_{\bar{i}} = \v{g}_i - \v{d}_i, 
	\label{eq:maopf_theta.power.int} \\
	&
	\quad \v{B}_{\bar{i},i} \v{\delta}_i + \v{B}_{\bar{i},\bar{i}} \v{\delta}_{\bar{i}} + \sum_{j \in \nbd(i)} \v{B}_{\bar{i},\bar{j}} \v{\delta}_{\bar{j}}
	= - \v{d}_{\bar{i}}, 
	\label{eq:maopf_theta.power.bnd} \\
	&
	\quad - \v{f}_i \leq \v{H}_{i,i} \v{\delta}_i + \v{H}_{i,\bar{i}} \v{\delta}_{\bar{i}} \leq \v{f}_i,
	\label{eq:maopf_theta.line.int} \\
	&
	\quad - \v{f}_{\bar{i}} \leq \v{H}_{\bar{i},\bar{i}} \v{\delta}_{\bar{i}} + \sum_{j \in \nbd(i)} \v{H}_{\bar{i},\bar{j}} \v{\delta}_{\bar{j}} \leq \v{f}_{\bar{i}}, 
	\label{eq:maopf_theta.line.TL} \\	
	&
	\quad \underline{\v{g}}_i \leq \v{g}_i \leq \overline{\v{g}}_i, 
	\label{eq:maopf_theta.gen.int} \\
    & \quad \delta^{\textrm{ref}} = 0, \quad \forall \  i = 1 , \dots, N, \label{eq:maopf_theta.ref}
	\end{alignat}
\end{subequations}
where, as shown in Fig. \ref{fig:maopf}, decision variables include area $i$'s power generation $\v{g}_i$, internal and boundary phase angles $\v{\delta}_i$, $\v{\delta}_{\bar{i}}$. 
Both loads $\v{d}_i$ and $\v{d}_{\bar{i}}$ are constants.
The objective \eqref{eq:maopf_theta.obj} is to minimize the sum of all area's generation costs with coefficients $\v{Q}_i$, $\v{c}_i$.
The DC model's nodal power balance is divided into \eqref{eq:maopf_theta.power.int}-\eqref{eq:maopf_theta.power.bnd}.
Notation $\nbd(i)$ collects all adjacent areas of area $i$. The subscripts $i$, $\bar{i}$, and $\bar{j}$ reflect bus partitions. 
Constraints \eqref{eq:maopf_theta.line.int}-\eqref{eq:maopf_theta.line.TL} restrict internal and tie-line branch flow less than $\v{f}_i$ and $\v{f}_{\bar{i}}$. 
Here, we slightly abuse the notations to let the row indices $i$, $\bar{i}$ be the internal and tie-line of area $i$. And column indices $i$, $\bar{i}$, $\bar{j}$ of $\v{H}$ are still the bus partitions. 
The lower and upper bounds of generation capacities $\underline{\v{g}}_i$, $\overline{\v{g}}_i$ are summarized in \eqref{eq:maopf_theta.gen.int}.
We assume that there is no generator on the boundary of each area. This way, the coordinator does not have direct jurisdiction over each area's generator. Such an assumption is not limiting. One can always derive an equivalent power network in Fig. \ref{fig:maopf} even with the presence of boundary generators, \cf, \cite{guoCoordinatedMultiAreaEconomic2017} for a treatment.
Constraint \eqref{eq:maopf_theta.ref} artificially assigns a reference phase angle.

\vspace{-10pt}
\section{Case studies}
\label{sec:case_new}

\subsection{Benchmarks and simulation settings}

Network data were obtained from MATPOWER 7.1 \cite{Zimmerman2011MATPOWERSO}. We designed eight multi-area systems as shown in Tab. \ref{tab:cre_asr.lin}. 
Notation \code{3area14$+$30$\times$2} means such a system consists of three areas stitching together. And the area's networks are taken from MATPOWER 7.1's standard case files, \ie, \code{case14.m}, \code{case30.m}, and \code{case30.m}, respectively. The topologies and settings of all networks are detailed in Fig. 9-10 in Appendix \ref{apdx:topology}

\vspace{-5pt}
\subsection{Computation time comparisons under a cold start}
\label{subsec: case.cold_start}

Our first set of simulations compares the convergence of CRE under the proposed degeneracy handling method. The implementation is detailed in Algo. \ref{alg:LC-Pivoting}-\ref{alg:CRE_ARS}, where numerical tolerances are defined in Algo. \ref{alg:CRE_ARS}. A cold start, \ie, $\v{\theta}^0 = \v{0}$, is adopted for all benchmarks. And all generators have linear generation costs. 
We compare such improved CRE with two other classical distributed algorithms, \ie, ADMM from \cite[Algo. 2]{kargarianDistributedDecentralizedDC2018}, and Benders decomposition from \cite[Sec. 5.1]{birgeIntroductionStochasticProgramming2011}, also under a cold start. The averaged results, which have been run ten times, are shown in Fig. \ref{fig:time.lin.cold_compare}. Detailed simulation settings and additional convergence curves can be found in Appendix \ref{apdx:simul}. 
The convergence superiority of CRE over the existing approaches under nondegenerate cases has been verified in \cite{guoCoordinatedMultiAreaEconomic2017,guoRobustTieLineScheduling2018}. 
With the proposed technique, the improved CRE method can ensure convergence and has a comparable total computation time cost with state-of-the-art techniques, even if degeneracy exists. The phenomena become more obvious, especially when the system scales. This is due to the much fewer iterations CRE needs to converge to the exact optimal solutions than other methods.
\begin{figure}[htbp]
    \vspace{-10pt}
    \centering
    \includegraphics[width=0.4\textwidth]{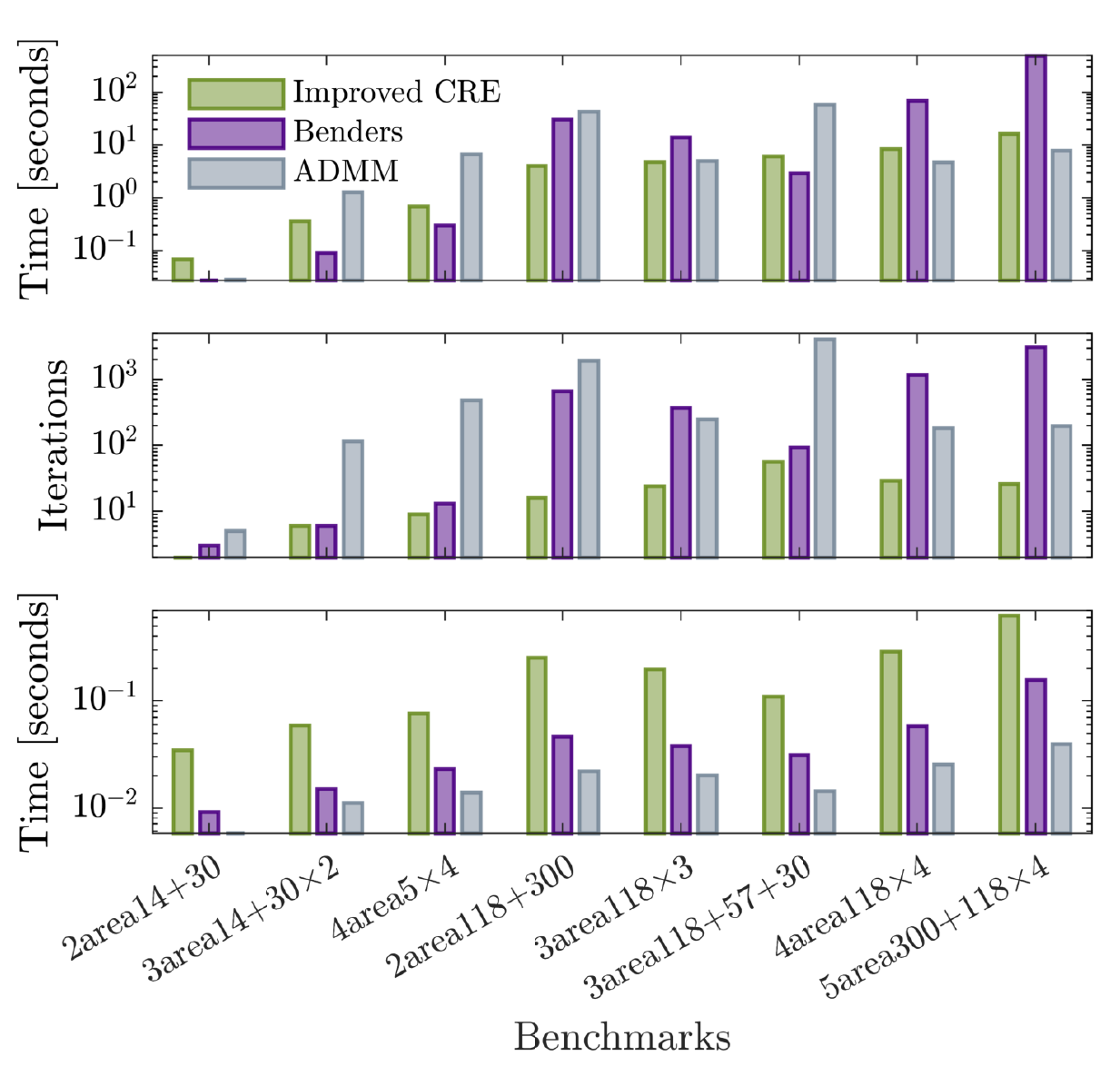}      
    \vspace{-10pt}           
	\caption{Total time (top), iterations (middle), per iteration time (bottom, [Total time] / [Total iterations]) comparisons among CRE, Benders, and ADMM.}
    \vspace{-10pt}
    \label{fig:time.lin.cold_compare}
\end{figure}

\vspace{-5pt}
\subsection{Analysis of degeneracy handling of each benchmark}

To further analyze the per iteration time cost due to degeneracy handling, the improved CRE in Algo. \ref{alg:CRE_ARS} is subdivided into CRE solving and degeneracy handling modules based on the following rules.
\begin{enumerate}[leftmargin=*]
    \item \emph{CRE solving}: The total time of Algo. \ref{alg:CRE_ARS} is executed, except steps \ref{alg:CRE_ARS.LC4}-\ref{alg:CRE_ARS.LC5} of Algo. \ref{alg:CRE_ARS} where Algo. \ref{alg:LC-Pivoting} has been invoked,
    \item \emph{Degeneracy handling}: Total time is recorded when steps \ref{alg:LC-Pivoting.deg.lcp}-\ref{alg:LC-Pivoting.deg.basis} of Algo. \ref{alg:LC-Pivoting} is invoked, and extra time is introduced by the nonunique basis in step \ref{alg:LC-Pivoting.map} of Algo. \ref{alg:LC-Pivoting}, step \ref{alg:CRE_ARS.co.opt} of Algo. \ref{alg:CRE_ARS} has also been taken into account.
\end{enumerate}

\begin{figure}[htbp]
    \vspace{-10pt}
    \centering
    \includegraphics[width=0.4\textwidth]{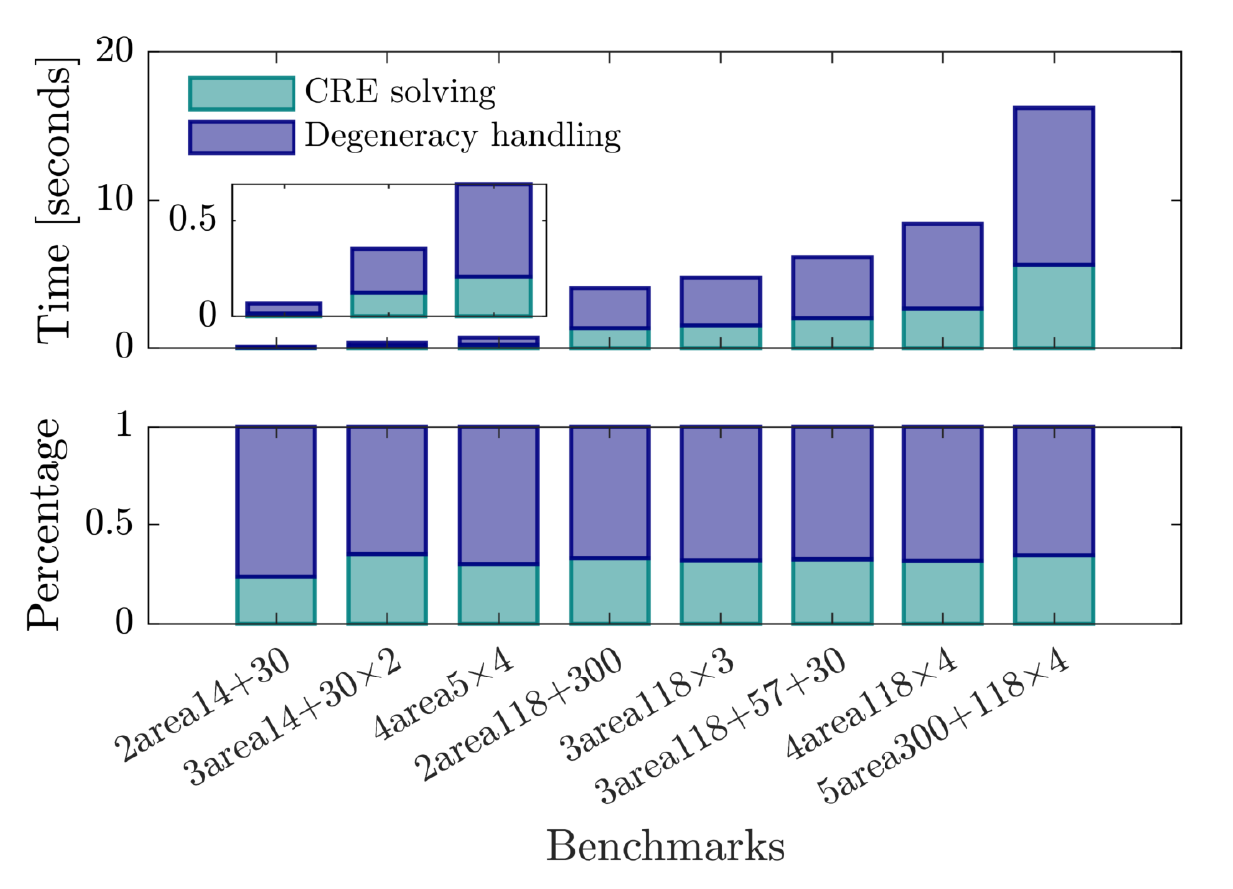}  
    \vspace{-10pt}           
	\caption{Time comparison of CRE solving and degeneracy handling. The ``Percentage" subplot implies the percentage of time about CRE solving and Degeneracy handling regarding the total computation time.}
    \label{fig:time.lin.cold_cre}
\end{figure}

As shown in Fig. \ref{fig:time.lin.cold_cre}, degeneracy handling takes up approximately 65\%-70\% of the total solving time during the process. Such handling can robustly ensure the convergence of the improved CRE in finite time, regardless of the system's complexity. 
The dominant time consumption of degeneracy handling is from the enumeration of the candidate basis for the degenerate vertex. We should point out that most candidates are invalid, as they do not satisfy the rank test \eqref{eq:mpLCP.rank}. 
\vspace{-5pt}

More specifically, we record the following three indices to reflect the degree of degeneracy in each case:
Index [Vertex] records the number of maximum vertices and total iterations when local problems have multiple vertices, shown in steps \ref{alg:LC-Pivoting.deg.unique2} and \ref{alg:LC-Pivoting.deg.nonunique2}.
Similarly, [Basis] records the number of maximum bases and total iterations when local problems have multiple bases, which can be obtained from step \ref{alg:LC-Pivoting.deg.basis}.
Besides, the [Iter.] shows the total iterations until reaching the exact convergence in steps \ref{alg:CRE_ARS.co.return1}-\ref{alg:CRE_ARS.co.return2} of Algo. \ref{alg:CRE_ARS}.
The results are shown in Tab. \ref{tab:cre_asr.lin}. 
In our simulations, the number of vertices equals the number of bases. But this may not always be the case. More than one basis for a vertex may be identified by enumerating the indices of degenerate components.
For \code{2area14$+$30} and \code{3area14$+$30$\times$2} cases, the vertex and basis are unique during the entire iterations. Such situations imply there is a unique $\CR$, and it is easy to obtain regardless of degeneracy. The proposed degeneracy handling process might be less efficient in these cases. The lexicographic selection rule suggested in \cite{jonesLexicographicPerturbationMultiparametric2007} might be a better option.
However, the significance of the proposed method becomes visible with the growing system complexity in the remaining cases. Multiple degenerate vertices/basis are identified during iterations, where each basis can generate a $\CR$ for coordinated optimization \eqref{eq:cre.coordination}.
As a result, the proposed degeneracy handling approach ensures a stable, fast finite convergence property and overall efficiency of the improved CRE method.
\begin{table}[htbp!]
    \centering
    \vspace{-10pt}
    \caption{Convergence of CRE under degeneracy handling method.}
    \label{tab:cre_asr.lin}
    \begin{tabularx}{0.4\textwidth}{l | CCC}
        \toprule   
        Networks  &  Vertex  &  Basis  &  Iter.  \\
        \midrule
        2area14$+$30 
            & 2/0 & 2/0 & 2 \\
        3area14$+$30$\times$2 
            & 3/0 & 3/0 & 6 \\ 
        4area5$\times$4 
            & 5/1 & 5/1 & 9 \\ 
        2area118$+$300
            & 3/14 & 3/14 & 16 \\  
        3area118$\times$3     
            & 4/1 & 4/1 & 24 \\               
        3area118$+$57$+$30    
            & 13/7 & 13/7 & 56 \\  
        4area118$\times$4 
            & 5/4 & 5/4 & 29 \\                    
        5area300$+$118$\times$4
            & 9/23 & 9/23 & 26 \\      
        \bottomrule       
    \end{tabularx}
\end{table}
\vspace{-15pt}
\subsection{Benchmark tests under random initial system states}

\begin{figure*}[htbp]
    \centering
    \vspace{-20pt}
    \includegraphics[width=0.9\textwidth]{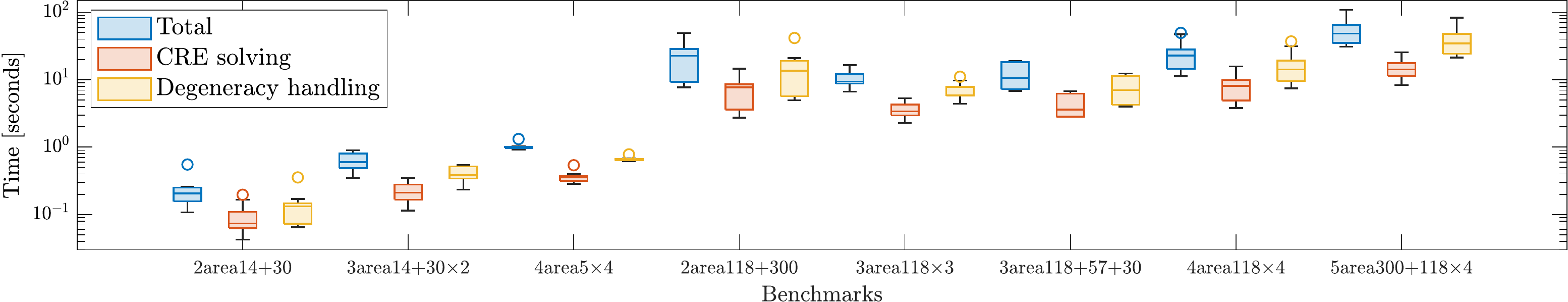}
    \vspace{-10pt}          
	\caption{Boxchart time comparison of CRE solving and degeneracy handling under ten random initial states.}
    \label{fig:time.lin.rand}    
    \vspace{-10pt}
\end{figure*}

Finally, the improved CRE’s benchmark performance is verified under various initial states.
Specifically, we randomly draw ten initial start points from a uniform distribution $\Ucal(-\v{\pi}, \v{\pi})$ for each benchmark. 
From Fig. \ref{fig:time.lin.rand}, the total solving time fluctuates within a small range for each case's random initial states. This represents that the total iterations are also similar for each benchmark. 
Moreover, the proposed degeneracy handling method ensures the improved CRE with a stable performance by requiring reasonable extra time for processing degeneracy. 
Typically, the conditions that trigger degeneracies for the tie-line scheduling system mainly lie in twofold:
\begin{enumerate}[leftmargin=*]
    \item Nonunique generation schedule (dual degeneracy): it is due to a set of generators has the same marginal costs under a certain tie-line schedule,
    \item Nonunique locational marginal prices (LMP) (primal degeneracy): 
    A fixed tie-line schedule can be viewed as a flexible load during iterations. 
    Since LMP is a piecewise curve, the switch between the segments under different load levels implies the change of marginal generators.
    In the transition state, the LMP might be discontinuous and has nonunique prices, which is a sign of degeneracy.
\end{enumerate}

As shown in Fig. \ref{fig:time.lin.rand}, the improved CRE method can be smoothly applied to various system topologies and initial states by integrating the proposed degeneracy handling process. 
This indicates the improved CRE has significant potential to adapt to a wide range of practical problems such as coordination of DERs \cite{wangCloudComputingLocal2020} or joint transmission \& distribution networks \cite{linDecentralizedDynamicEconomic2018}. 
And the comprehensive handling of degeneracies under given parameters is also helpful to improve the performance of existing geometric-based mpLP/QP algorithms \cite{borrelliGeometricAlgorithmMultiparametric2003} or guide spot pricing when Lagrange multipliers are nonunique \cite{fengSpotPricingWhen2012}. 



\section{Conclusions}
\label{sec:conclusion}

This paper proposes a degeneracy handling method that leverages mpLCP formulation to identify and analyze degenerate situations in a unified view. 
The proposed method can find all full-dimensional critical regions containing the given parameter. This is achieved by an efficient enumeration to the basis of the complementary solution set's vertices.
A general CRE coordination process is also illustrated to show how to solve a distributed optimization problem.
An improved CRE is designed, which integrates the proposed degeneracy handling method, applies an adaptive stepsize to explore adjacent critical regions, and recovers feasibilities by cutting plane updates with projection.
The improved CRE shows comparable performance with state-of-art methods on eight tie-line scheduling benchmarks and outperforms when the system scales.
Moreover, with the proposed degeneracy handling technique, CRE can ensure fast finite convergence under various system conditions.
Such efficiency indicates applying the improved CRE to multiple power system applications is promising. 
And the effectiveness of the proposed degeneracy handling shows potential in analyzing practical degenerated problems and complementing mpP algorithm design.




\section*{Acknowledgment}
The authors appreciate the constructive discussions with Professor Subhonmesh Bose from the University of Illinois Urbana-Champaign.


\bibliographystyle{IEEEtran}
\bibliography{IEEEabrv,ref_jnl_sorted}


\appendix

\subsection{Recover linear complementary problem solution from commercial solvers}
\label{apdx:solver}

Lemke's method is not quite an efficient algorithm. One may prefer using commercial solvers (Cplex, Mosek, Gurobi, \etc.) to solve large-scale problems efficiently. There are two dominant types of algorithms for modern solvers. One is the Simplex method for the LP problem ($\v{H} = 0$), and the other is the interior point method (IPM) for other general convex problems.

We aim to obtain a vertex solution from the optimal primal-dual set of problem \eqref{eq:mpLCP_ini}. For the Simplex method already has an optimal vertex solution when converged. 
Let $\{ \Bcal_x, \Ncal_x \}$, $\{ \Bcal_s, \Ncal_s \}$ be the basic and nonbasic indices to the primal and slack variables $[\v{x}^\T, \v{s}^\T ]^\T$ at final iteration. Then the KKT condition \eqref{eq:mpLCP_mtx} when $\v{H} = \v{0}$ can be rewritten as
\begin{align}
    \begin{bmatrix}
        \v{\mu}_{\Bcal_x} \\
        \v{\mu}_{\Ncal_x} \\
        \v{s}_{\Bcal_s} \\
        \v{s}_{\Ncal_s} 
    \end{bmatrix} = & \begin{bmatrix}
        \v{0} & \v{0} & \v{A}^\T_{\Bcal_s, \Bcal_x} & \v{A}^\T_{\Ncal_s, \Bcal_x} \\
        \v{0} & \v{0} & \v{A}^\T_{\Bcal_s, \Ncal_x} & \v{A}^\T_{\Ncal_s, \Ncal_x} \\
        -\v{A}_{\Bcal_s, \Bcal_x} & -\v{A}_{\Bcal_s, \Ncal_x} & \v{0} & \v{0} \\ 
        -\v{A}_{\Ncal_s, \Bcal_x} & -\v{A}_{\Ncal_s, \Ncal_x} & \v{0} & \v{0} \\ 
    \end{bmatrix} \notag \\ 
    & \cdot \begin{bmatrix}
        \v{x}_{\Bcal_x} \\
        \v{x}_{\Ncal_x} \\
        \v{\lambda}_{\Bcal_s} \\
        \v{\lambda}_{\Ncal_s} \\
    \end{bmatrix} + \begin{bmatrix}
        \v{f}_{\Bcal_x} \\
        \v{f}_{\Ncal_x} \\
        \v{b}_{\Bcal_s} \\
        \v{b}_{\Ncal_s}
    \end{bmatrix} + \begin{bmatrix}
        \v{0} \\
        \v{0} \\
        \v{C}_{\Bcal_s} \\
        \v{C}_{\Ncal_s}
    \end{bmatrix} \v{\theta}.
    \label{eq:Simplex.mtx.partition}
\end{align}

Based on \eqref{eq:Simplex.mtx.partition}, we can directly know a basic variable for mpLCP is $\v{y}_{\hat{\Bcal}} = [ \v{x}_{\Bcal_x}^\T, \v{\lambda}_{\Bcal_s}^\T, \v{\mu}_{\Ncal_x}^\T, \v{s}_{\Ncal_s}^\T ]^\T$. This is due to (i) the Simplex's basic variables $[\v{x}_{\Bcal_x}^\T, \v{s}_{\Bcal_s}^\T] \geq \v{0}$, and they are in the basis, (ii) the complementary rule of mpLCP implies $[\v{\mu}_{\Bcal_x}^\T, \v{\lambda}_{\Bcal_s}^\T]$ are not in the basis. Note the basis $\hat{\Bcal}$ recovered from the Simplex method's solutions may not be the same as $\bar{\Bcal}$ from Lemke's method, as they may have different pivot rules. Nevertheless, they all reach a vertex to the optimal set. When there is no degeneracy, then $\hat{\Bcal} = \bar{\Bcal}$ must always hold.

Recovering the basis is not direct for the IPM as it may converge at an interior solution to the optimal set. Rather than adopting pivot operations, IPM traverses the interior of the feasible regions. Let $(\v{x}^\star, \v{\mu}^\star, \v{s}^\star, \v{\lambda}^\star)$ be the optimal solution pair of \eqref{eq:mpLCP_ini} from IPM. Here, we apply a simple check based on the complementary property to recover a basic solution, which is given by
\begin{gather}
    { 
    \begin{gathered}
        \v{y}_{\hat{\Bcal}}^\star = [ \max(\v{x}^\star, \v{\mu}^\star)^\T, \ \max(\v{s}^\star, \v{\lambda}^\star)^\T ]^\T, \\
        \v{y}_{\hat{\Ncal}}^\star = [ \min(\v{x}^\star, \v{\mu}^\star)^\T, \ \min(\v{s}^\star, \v{\lambda}^\star)^\T ]^\T,    
        \label{eq:solver.IPM}
    \end{gathered}
    }
\end{gather}
where notations $\max$, $\min$ represent componentwise maximum / minimum. And $(\v{y}_{\hat{\Bcal}}^\star, \v{y}_{\hat{\Ncal}}^\star)$ represents the solution partition. Under nondegenerated conditions, we have $\v{y}_{\hat{\Bcal}}^\star > 0$, $\v{y}_{\hat{\Ncal}}^\star = 0$. When there are elements $x_i^\star = \mu_i^\star = 0$ or $s_i^\star = \lambda_i^\star = 0$ as a result of degeneracy, we randomly select one to be the basic variable. 
The results from the Simplex method can also apply \eqref{eq:solver.IPM} to recover the basic solution if the solver does not return any basis information.
Note that the coefficients in \eqref{eq:mpLCP.Lemke.ini.coef} are unknown if we apply the Simplex method or IPM. We need to invert under $\hat{\Bcal}$ to generate them.

\subsection{Testing benchmark topologies}
\label{apdx:topology}

Unless specified, all internal lines are taking default capacities from MATPOWER 7.1. If there is no capacity given, then we will set it as 800 MW.

\begin{figure}[htbp]
    \centering 
    \vspace{-0.15in}
    \subfloat[Two area 14$+$30 network]{
        \includegraphics[width=0.4\textwidth]{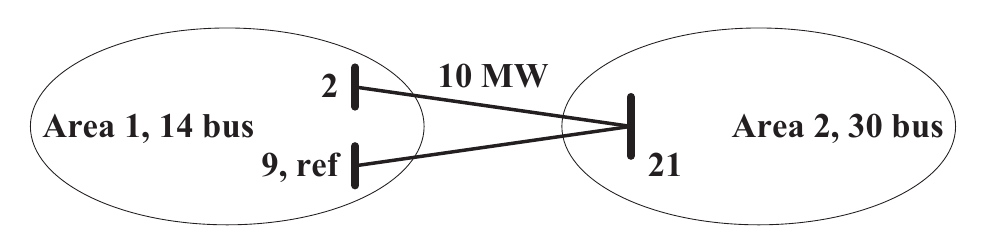}
    }  
    \\
    \subfloat[Three area 14$+$30$\times$2 network]{
        \includegraphics[width=0.4\textwidth]{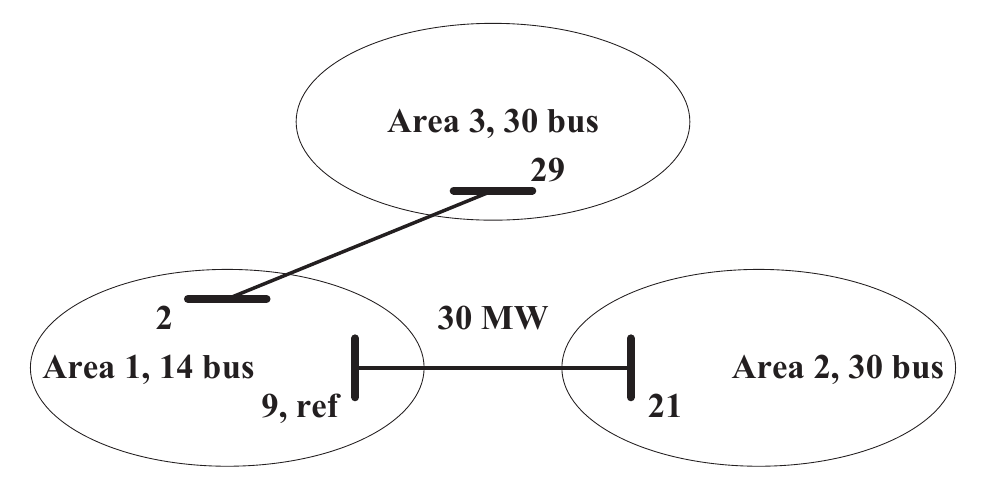}
    }     
    \\
    \subfloat[Four area 5$\times$4 network]{
        \includegraphics[width=0.4\textwidth]{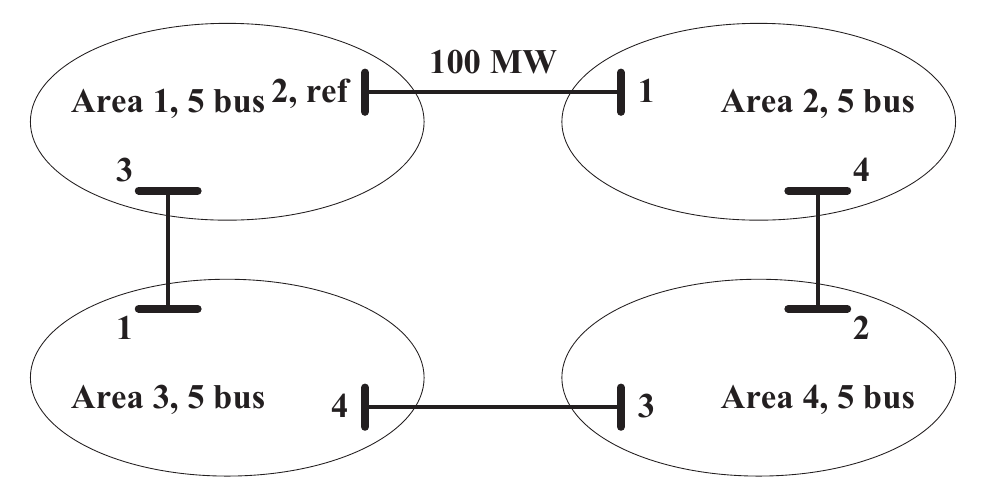} 
    }  
    \\
    \subfloat[Two area 118$+$300 network]{
        \includegraphics[width=0.4\textwidth]{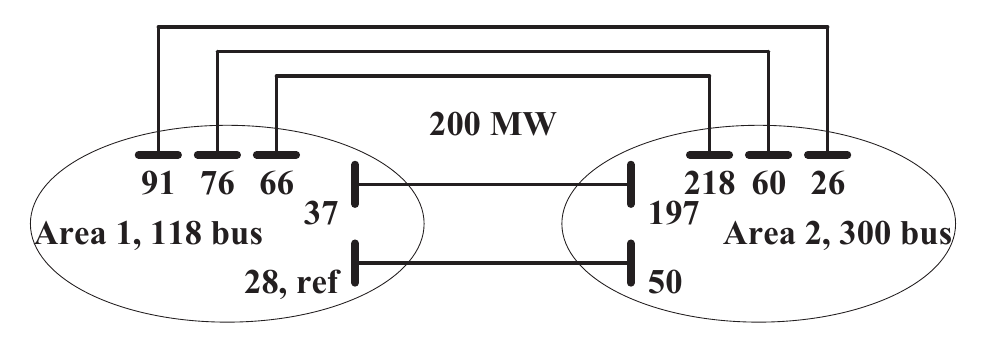}
    }      
    \\
    \subfloat[Three area 118$\times$3 network]{
        \includegraphics[width=0.4\textwidth]{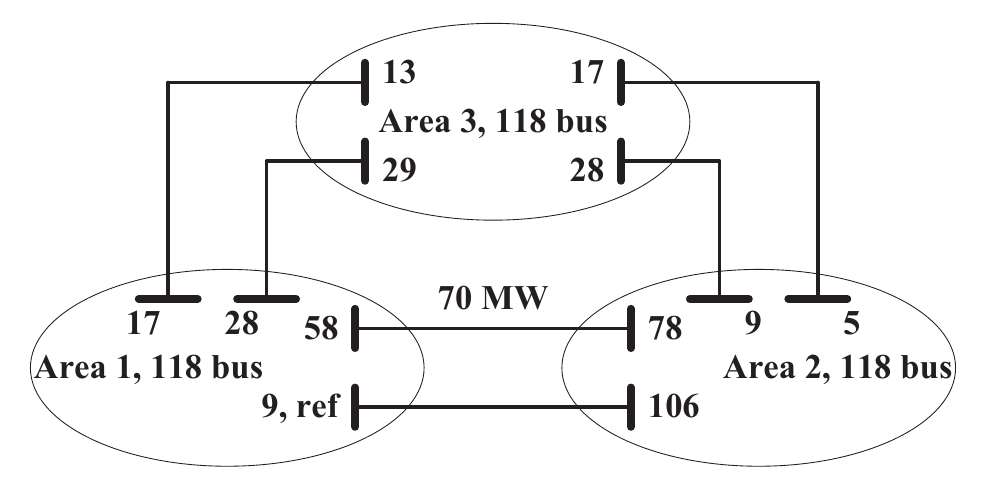}
    }            
    \caption{ Testing benchmarks for multi-area network systems. }
    \label{fig:benchmark1}
\end{figure}

\newpage

\begin{figure}[htbp]
    \centering 
    \subfloat[Three area 118$+$57$+$30 network]{
        \includegraphics[width=0.4\textwidth]{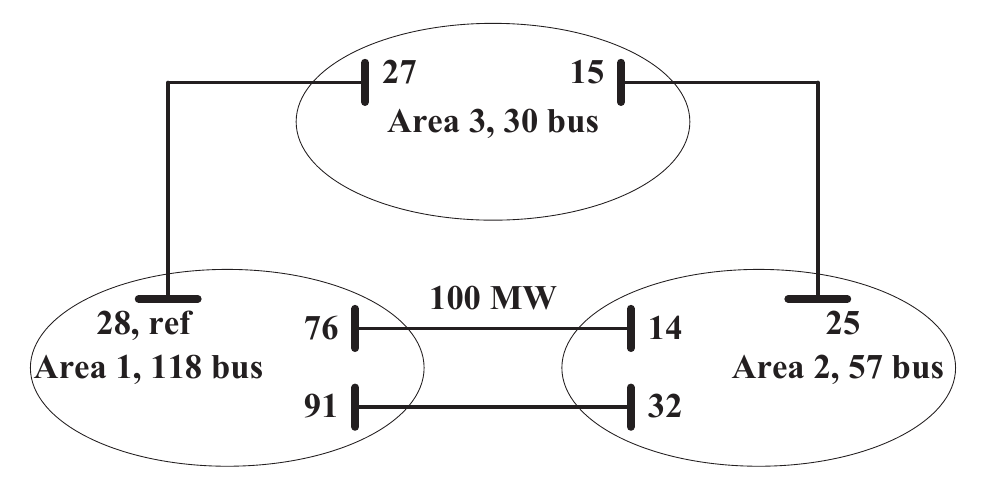}
    } 
    \\
    \subfloat[Four area 118$\times$4 network]{
        \includegraphics[width=0.4\textwidth]{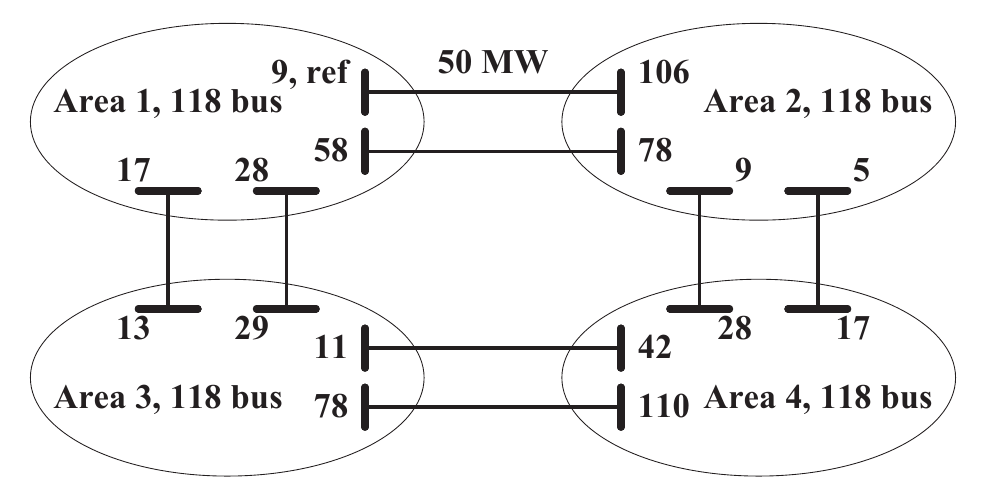}
    }
    \\
    \subfloat[Five area 300$+$118$\times$4 network]{
        \includegraphics[width=0.4\textwidth]       
        {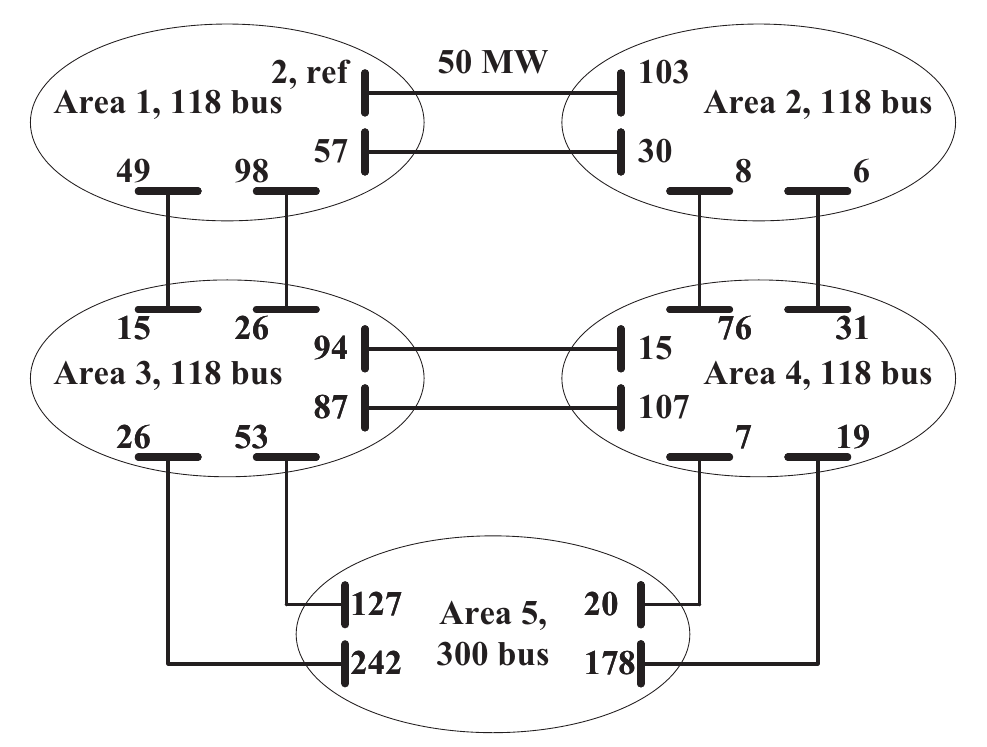}
    }
    \caption{ Testing benchmarks for multi-area network systems, continue. }
    \label{fig:benchmark2}
\end{figure}

\subsection{Simulation setting and supplymentary results}
\label{apdx:simul}

We used MATLAB 2021a with Gurobi v9.5.2 to conduct all the simulations on a laptop. 
For the improve CRE, the problems \eqref{eq:cre.coordination}, \eqref{eq:CRE.LSQ}, and \eqref{eq:mpP_fea} were handled by Gurobi. 
Besides, we also used Gurobi to optimize \eqref{eq:mpLCP_ini} under $\v{\theta}^k$ instead of solving a large-scale mpLCP \eqref{eq:mpLCP_fin} directly.
Lemke's method solved the auxiliary LCP \eqref{eq:LCP_aux} with the lexicographic minimum row selection technique \cite[Ch. 2]{Murty1988-lx}.
And MPT 3.0\cite{MPT3}'s \code{P.minVRep()} was adopted to search for the minimal optimal vertices in step \ref{alg:LC-Pivoting.deg.nonunique2} of Algo. \ref{alg:LC-Pivoting}. 
Generally, partitioning boundary phase angles' $\v{\Theta}^\star$ into $\CR$ usually results in many tiny critical regions. To partially relieve this phenomenon, we equivalently enlarged all the $\CR$ by scaling the parameters' coefficients in \eqref{eq:probP.local}-\eqref{eq:probP.couple} as
\begin{gather}
    \tilde{\v{C}}_i = 0.01\v{C}_i, \ \tilde{\v{D}}_i = 0.01\v{D}_i, \ i = 1, \ldots, N,
    \label{eq:scale}
\end{gather}
where $\tilde{\v{C}}_i$ and $\tilde{\v{D}}_i$ have been substituted into the original problem. Note that the optimal $\JJ(\v{\theta}^\star)$ will not change, whereas $\v{\theta}^\star$ and $\CR$ are enlarged accordingly.

\begin{figure}[htbp]
    \centering 
    \subfloat[Two area 14$+$30 network]{
        \includegraphics[width=0.475\textwidth]{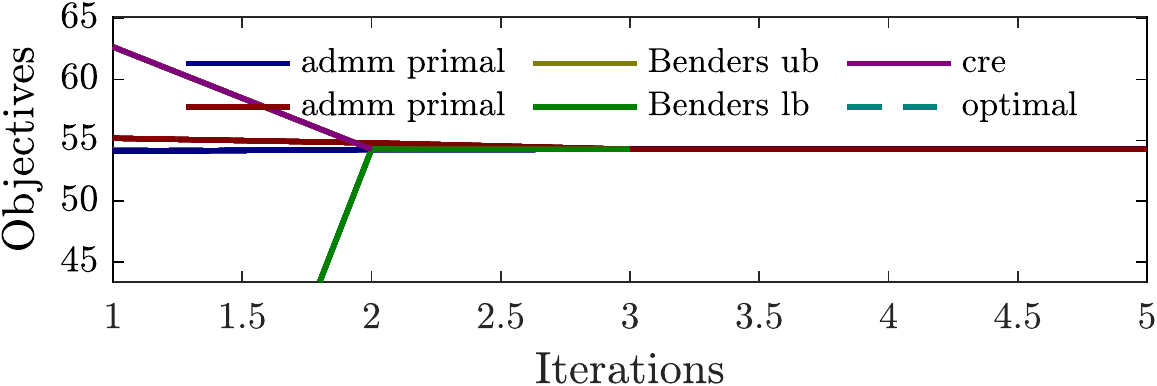}
    }  
    \\
    \subfloat[Three area 14$+$30$\times$2 network]{
        \includegraphics[width=0.475\textwidth]{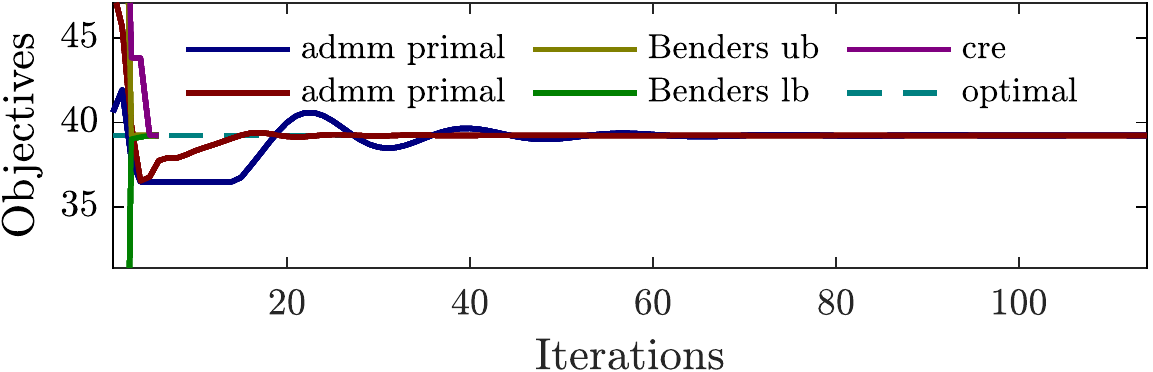}
    }     
    \\
    \subfloat[Four area 5$\times$4 network]{
        \includegraphics[width=0.475\textwidth]{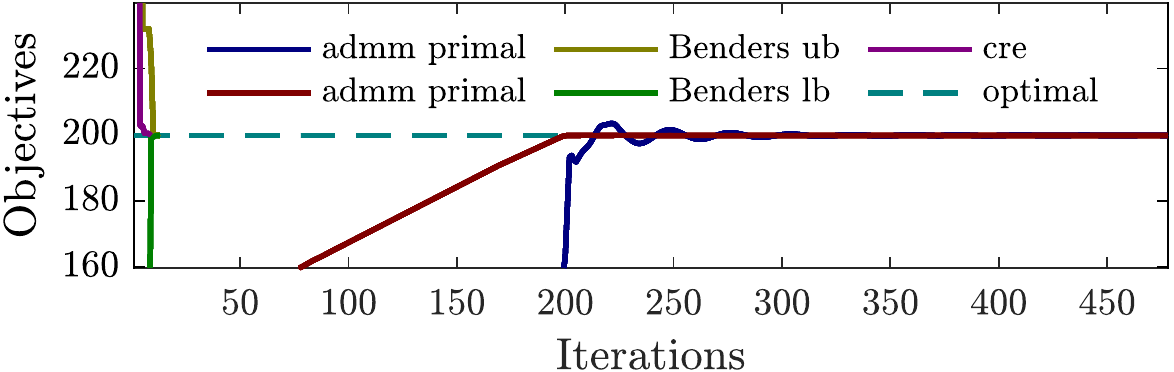}
    }  
    \\
    \subfloat[Two area 118$+$300 network]{
        \includegraphics[width=0.475\textwidth]{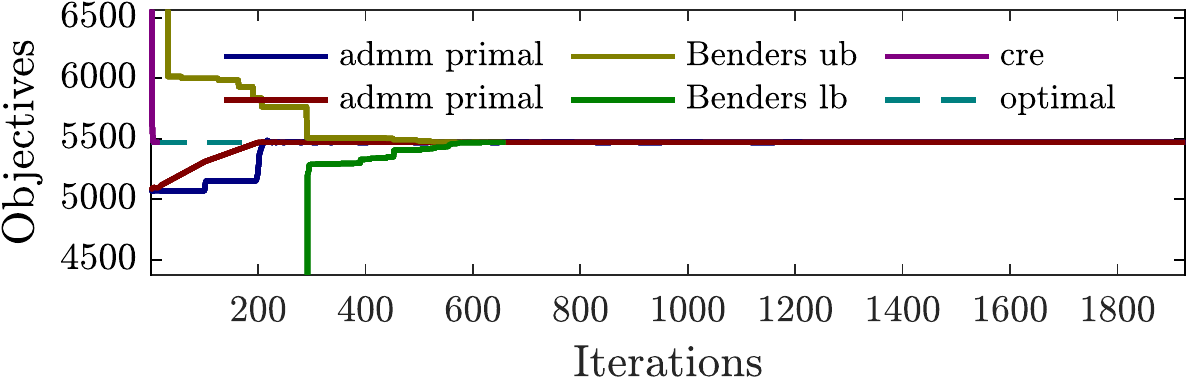}
    }   
    \caption{Cold start comparisons on testing benchmarks.}
    \label{fig:iter1}
\end{figure}

\begin{figure}[htbp]
    \centering 
    \subfloat[Three area 118$\times$3 network]{
        \includegraphics[width=0.475\textwidth]{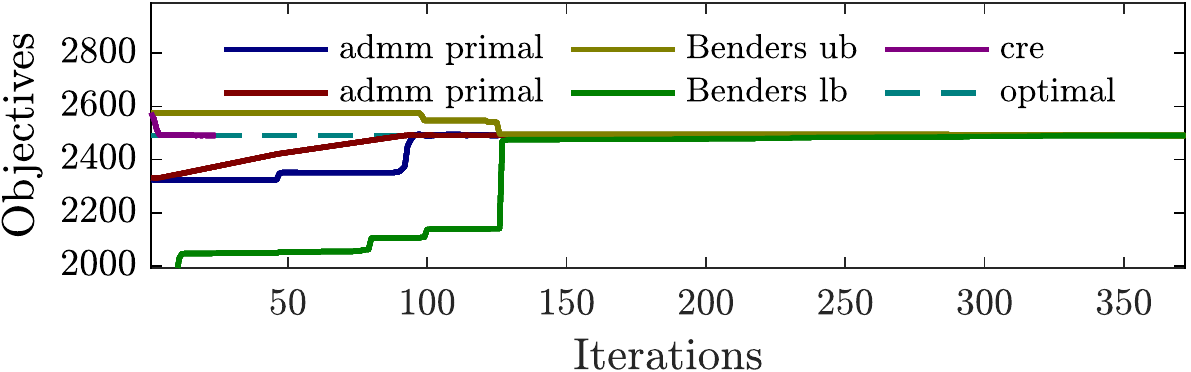}
    }     
    \\ 
    \subfloat[Three area 118$+$57$+$30 network]{
        \includegraphics[width=0.475\textwidth]{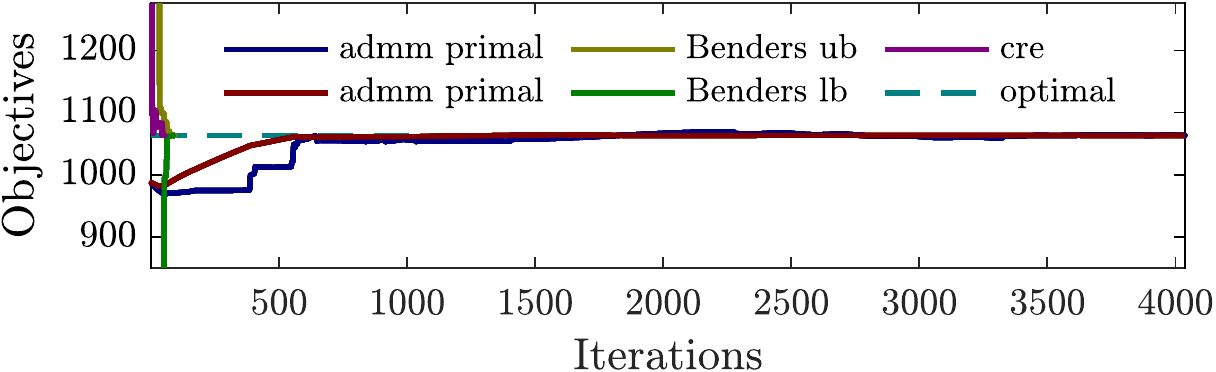}
    } 
    \\
    \subfloat[Four area 118$\times$4 network]{
        \includegraphics[width=0.475\textwidth]{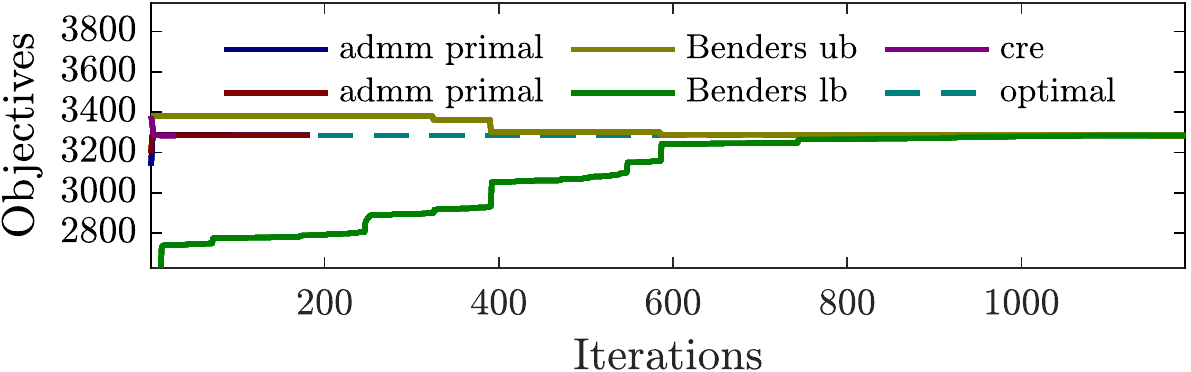}
    }
    \\
    \subfloat[Five area 300$+$118$\times$4 network]{
        \includegraphics[width=0.475\textwidth]{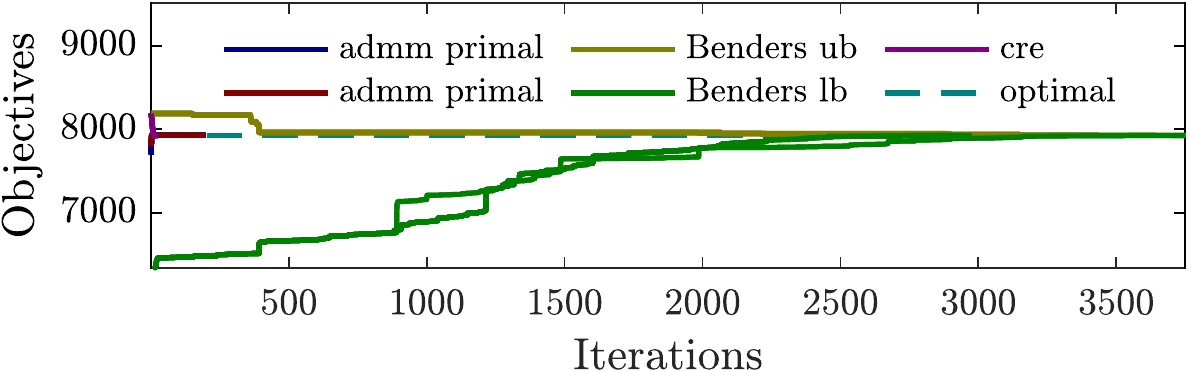}     
        \label{fig: nonconvergent} 
    }            
    \caption{Cold start comparisons on testing benchmarks, continue.}
    \label{fig:iter2}
\end{figure}

We use gradient norm for convergence measurement of CRE, \ie, $\lVert \v{v}^\star \rVert \leq 10^{-2}$, where $\v{v}^\star$ is calculated from \eqref{eq:CRE.LSQ}. 
For the ADMM method, we set the penalty factor equals to $0.1$. The relative optimality gap metric is given as
\begin{gather} 
    { 
    \begin{gathered}
        |f_{\Pcal}^k - f_{\Dcal}^k| + \sum_{i=1}^N \|\v{\theta}_i - \v{\theta} \|_1 \leq 10^{-3},
    \end{gathered}
    }
\end{gather}
where $f_{\Pcal}^k$, $f_{\Dcal}^k$ represents the primal and dual objective value at $k^{\textrm{th}}$ iteration. The second term measures the constraint violation after introducing copies $\v{\theta}_i$ of the global consensus boundary state $\v{\theta}$.
For Benders decomposition, we initialize the lower and upper bound of the objective function as $f_{\Lcal}^0=0$, $f_{\Ucal}^0=10^5$, respectively. The relative optimality gap metric for termination is given as $f_{\Ucal}^k - f_{\Lcal}^k \leq 10^{-3}$.
For the eight cases as shown in Fig. \ref{fig:benchmark1}-\ref{fig:benchmark2}, the convergence of iterations under a cold start is shown in Fig. \ref{fig:iter1}-\ref{fig:iter2}.

\newpage

As demonstrated in Fig. \ref{fig: nonconvergent}, the Benders method has unstable convergence issues for large-scale problems under ten repeated trials. This is mainly due to the solvers' accuracy for the solution, which affects the quality of the cutting plane generation. The convergence of ADMM on a higher accuracy is much more challenging, as the dual gradients are nearly vanishing when approaching optimal. Nevertheless, the improved CRE can always ensure a fast and stable convergence performance on the above benchmarks.


\end{document}